\documentclass[12pt]{amsart}
\usepackage{amsmath,amssymb,enumitem,hyperref,cite,graphicx}

\hypersetup{pdftex,colorlinks=true,allcolors=blue}
\usepackage[all]{hypcap}

\usepackage[foot]{amsaddr}
\usepackage[sc]{mathpazo}

\newcommand\ack{\subsection*{Acknowledgment}}

\DeclareMathAlphabet\mathsfbi{T1}{phv}{b}{it}

\numberwithin{equation}{section}

\newcommand\BV{\boldsymbol} 
\newcommand\BM{\mathsfbi} 

\newcommand\dif{\:\!\mathrm{d}}
\newcommand\deriv[2]{\frac{\mathrm{d} #1}{\mathrm{d} #2}}
\newcommand\parderiv[2]{\frac{\partial #1}{\partial #2}}
\newcommand\Dderiv[2]{\frac{\mathrm D #1}{\mathrm D #2}}
\newcommand\RR{\mathbb{R}}
\newcommand\coll{\mathcal C}
\newcommand\Mach{\mathit{Ma}}
\newcommand\Rey{\mathit{Re}}
\newcommand\Pran{\mathit{Pr}}
\newcommand\Knud{\mathit{Kn}}
\newcommand\mfp{l}

\DeclareMathOperator{\cL}{\mathcal{L}}
\DeclareMathOperator{\EE}{\mathbb{E}}
\DeclareMathOperator{\trace}{\mathrm{tr}}

\def\HS{{HS}}

\newcommand\myatop[2]{\genfrac{}{}{0pt}{}{#1}{#2}}

\begin{document}

\author[Rafail V. Abramov]{Rafail V. Abramov}

\address{Department of Mathematics, Statistics and Computer Science,
University of Illinois at Chicago, 851 S. Morgan st., Chicago, IL 60607}

\email{abramov@uic.edu}

\title[A molecular-kinetic hypothesis on the mechanics of compressible
  gas flow]{A molecular-kinetic hypothesis on the mechanics of
  compressible gas flow at low Mach numbers}

\begin{abstract}
In recent works, we proposed a theory of turbulence creation via the
second coefficient of the virial expansion (i.e.~the van der Waals
effect). This theory relies, in part, on the empirically observed
``equilibrated'' behavior of pressure in compressible flows at low
Mach numbers. However, a fundamental explanation for such a behavior
of pressure does not currently exist, because the conventional kinetic
theory leads instead to the adiabatic flow in the form of the usual
compressible Euler or Navier--Stokes equations.

To explain this behavior of pressure from the molecular-kinetic
perspective, in the current work we introduce a novel correction into
the pair correlation function in the closure of the
Bogoliubov--Born--Green--Kirkwood--Yvon hierarchy. This correction
matches the rate of change of the average distance between particles
to the macroscopic compression or expansion rate of the gas.
Remarkably, the novel correction introduces strong dissipation into
the pressure equation at low Mach numbers, which stabilizes the
pressure solution. At small scales, the novel dissipation effect
manifests as the second viscosity in the momentum equation, which
selectively suppresses the velocity divergence. As a result, the
second viscosity governs the linear instability which creates
turbulent dynamics, thereby setting the critical value of the Reynolds
number. The ratio of the second and shear viscosities, together with
the critical value of the Reynolds number, are proportional to the
reciprocal of the packing fraction.
\end{abstract}

\maketitle

\section{Introduction}
\label{sec:intro}

It is known through observations that, at relatively slow speeds (or
low Mach numbers), the pressure in the atmospheric air flow is mostly
equilibrated, whereas the density and temperature may vary
considerably. The phenomenon of convection is one of the consequences
of such a behavior --- indeed, when an air parcel warms up, it
expands, while its pressure remains the same. In turn, the expansion
leads to a lower density than the ambient air, so that the reduced
gravity force no longer balances the pressure gradient, which causes
positive buoyancy. Remarkably, this pressure stabilization is not
caused by the momentum viscosity or the heat conduction effects --- in
fact, the scale of a typical convection pattern far exceeds the
viscous scale at normal conditions (that is, the flow has a high
Reynolds number).

Moreover, in our recent works
\cite{Abr22,Abr23,Abr24,Abr26,Abr27,Abr25} we proposed a new model of
turbulence via the van der Waals effect in a compressible gas, where
the pressure variable was either set to a constant
\cite{Abr22,Abr23,Abr24,Abr26,Abr27}, or its equation was artificially
chosen to induce a linear damping in the velocity divergence
\cite{Abr25}.  In all studied cases, turbulent dynamics emerged
spontaneously from an initially laminar flow, just as observed in
nature and experiments. Therefore, the pressure stabilization at slow
speeds and in the near absence of viscous effects is consistent with
the presence of both convection and turbulent dynamics.

Yet, we still lack a fundamental understanding of such a pressure
behavior in the context of kinetic theory. In the absence of shear
viscosity (that is, at the infinite Reynolds number), the standard
equations for a compressible gas are the compressible Euler equations.
Remarkably, the compressible Euler equations fail to describe such a
stabilized behavior of their pressure variable at low Mach
numbers. Instead, they exhibit a directly opposite thermodynamic
behavior --- according to the Euler equations, the gas compresses when
its temperature increases, and expands, when it decreases.

To see how this happens, let us look at the compressible Euler
equations (see, for example, Section 2.1 of \cite{Gols}), expressed in
the density $\rho$, velocity $\BV u$ and pressure $p$ variables:
\begin{equation}
\label{eq:Euler}
\Dderiv\rho t+\rho\nabla\cdot\BV u=0,\qquad\rho\Dderiv{\BV u}t+\nabla
p=\BV 0,\qquad\Dderiv pt+\gamma p\nabla\cdot\BV u=0.
\end{equation}
Above, $\gamma>1$ is the adiabatic index (e.g. $\gamma=5/3$ for
monatomic gases), and
\begin{equation}
\Dderiv ft\equiv\parderiv ft+\BV u\cdot\nabla f
\end{equation}
is the advective derivative. First, we show that the entropy
\begin{equation}
\label{eq:entropy}
S=\frac p{\rho^\gamma}
\end{equation}
is preserved along stream lines (that is, the flow is adiabatic).
Indeed,
\begin{equation}
\Dderiv St=\Dderiv{}t\left(\frac p{\rho^\gamma}\right)=\frac 1{\rho
  ^\gamma}\Dderiv pt-\frac {\gamma p}{\rho^{\gamma+1}}\Dderiv\rho t=-
\frac{\gamma p}{\rho^\gamma}\nabla\cdot\BV u+\frac{\gamma p}{\rho^
  \gamma}\nabla\cdot\BV u=0.
\end{equation}
Next, we recall the equation of state of a dilute gas \cite{ChaCow},
\begin{equation}
\label{eq:state}
p=\rho\theta=\rho RT,
\end{equation}
where $T$ is the temperature, $R$ is the gas constant, and $\theta=RT$
is the kinetic temperature, which has the units of squared velocity.
With \eqref{eq:state}, the entropy \eqref{eq:entropy} is expressed via
\begin{equation}
S=\frac p{\rho^\gamma}=\frac{\rho RT}{\rho^\gamma}
=\frac{RT}{\rho^{\gamma-1}}.
\end{equation}
Since the quotient above is preserved along stream lines, the
temperature $T$ and density~$\rho$ increase and decrease
simultaneously in solutions of the Euler equations \eqref{eq:Euler}.

The compressible Navier--Stokes equations (refer, for instance, to
Section 2.2 of \cite{Gols}) are obtained from the compressible Euler
equations \eqref{eq:Euler} by adding the thermodynamically
irreversible effects of the momentum viscosity and heat conduction,
respectively, into the momentum and pressure transport
equations. However, at high Reynolds numbers those effects are small
compared to the advection, and the qualitative thermodynamic behavior
of density and temperature remains the same as for the Euler equations
\eqref{eq:Euler}.

Notably, this contradiction between reality and the solutions of the
Euler or Navier--Stokes equations manifests only at low speeds. At
high speeds, dilute gases behave as predicted by the Euler or
Navier--Stokes equations --- namely, they compress when heated and
expand when cooled, subsequently producing various adiabatic effects
such as the acoustic waves, shock transitions and Prandtl--Meyer
expansions.

Due to the above inconsistency, in practice the behavior of gases at
low Mach numbers is usually modeled via the incompressible Euler
equations (see, e.g.~Section 2.4 of \cite{Gols}):
\begin{equation}
\label{eq:incompressible_Euler}
\rho_0\Dderiv{\BV u}t+\nabla p=\BV 0,\qquad\nabla\cdot\BV u=0.
\end{equation}
Here, the density is preserved along the stream lines due to the
divergence-free velocity field, and can thereby be set to a constant
$\rho_0$. The pressure $p$ is no longer a thermodynamic variable, and
is instead chosen artificially to enforce the divergence-free velocity
condition. The incompressible Navier--Stokes equations are obtained by
adding a viscous dissipation into the right-hand side of the momentum
equation in \eqref{eq:incompressible_Euler}. While the compressible
Euler equations in \eqref{eq:Euler} can be derived from kinetic theory
\cite{Gols}, the incompressible Euler equations in
\eqref{eq:incompressible_Euler} are empirical --- namely, the
divergence-free velocity condition is imposed from observations of
behavior of gases at low Mach numbers, as well as liquids. Neither of
the two systems describes convection --- gases compress when heated in
the former, while the density of a gas is constant in the latter.

Since the Euler pressure equation in \eqref{eq:Euler} fails to model
thermodynamic properties of the flow in the low Mach number regime, a
na\"ive suggestion would be to correct it, in an appropriate fashion,
so that it adheres to the observed behavior of the gas. However, the
problem lies much deeper than it seems at a first sight, because the
entire set of the Euler equations \eqref{eq:Euler} is derived from the
single Boltzmann equation \cite{ChaCow,CerIllPul,Gra} for the velocity
distribution function, by computing the transport equations for its
velocity moments of appropriate order \cite{Gra,Abr17}.  Therefore, it
is impossible to ``correct'' the pressure equation separately, as it
would disconnect the latter from the Boltzmann equation.

This issue is further exacerbated by the fact that the possibility of
``correcting'' the Boltzmann equation also seems to be rather distant,
because it is, in turn, derived from the
Bogoliubov--Born--Green--Kirkwood--Yvon (BBGKY) hierarchy
\cite{Bog,BorGre,Kir,Yvo} of the corresponding Liouville equation for
the full multiparticle system. The Liouville equation itself is
ironclad; the only variable part of it is the intermolecular
potential. Thus, the entire chain of reasoning, starting from the
Liouville equation and ending at the Euler equations, is seemingly
immutable.  In particular, the inertial
\cite{Abr22,Abr23,Abr24,Abr26,Abr27} and weakly compressible
\cite{Abr25} pressure regimes, which lead to spontaneously developing
turbulent flows in our recent works, were introduced completely
empirically, solely by appealing to the natural, observable behavior
of real gases. The molecular-kinetic mechanism of the pressure
behavior at the low Mach numbers remained unknown.

\subsection{The new results in the current work}

Upon a close examination of the chain of reasoning, which leads from
the Liouville equation to the Euler equations, we concluded that the
only mutable part of it is the closure for the pair correlation
function in the BBGKY hierarchy. The current work is an attempt to
explain the behavior of dilute gases at low Mach numbers by
introducing a suitable correction to the pair correlation function.
The key highlights of this work are the following:
\begin{enumerate}[label=\arabic*.]
\item In a simplified, synthetic flow setting with a constant density
  and pressure, we use a direct calculation to find that the rate of
  change of the average distance in pairs of gas particles is
  proportional to the divergence of the flow velocity, that
  is,~$\nabla\cdot\BV u$.  At the same time, the infinitesimal
  generator of the two-particle distribution with the standard pair
  correlation function, derived from the Gibbs equilibrium state,
  yields zero rate of change irrespectively of the value of
  $\nabla\cdot\BV u$. It means that the standard pair correlation
  function fails to match the rate of change of the average distance
  in pairs of particles to the compression or expansion rate of the
  gas. This is explained in Section~\ref{sec:mean_distance}.
\item To hypothesize a suitable correction to the pair correlation
  function, we examine a pair of particles, which are distributed
  canonically, but have different average velocities, which depend on
  particle locations. We find that the requisite correction is related
  to the difference in particle velocities; when we incorporate it
  into the pair correlation function, the rate of change of the
  average pair distance becomes the exact match to the result of the
  direct calculation.  This is done in Section~\ref{sec:correction}.
\item Using the corrected pair correlation function in the collision
  integral, we recompute the transport equations for the density,
  momentum and pressure of the gas flow. It turns out that the novel
  correction does not affect the equations for the mass and momentum
  transport; only the pressure transport equation is affected. We find
  this in Section~\ref{sec:novel_equations}.
\item The novel term in the pressure equation involves the divergence
  of the velocity difference in a pair of particles. It is impossible
  to describe such quantity precisely in the context of the
  single-particle velocity moments, and thus we have to resort to a
  phenomenological closure. To achieve a closure, we take advantage of
  a running time average, which separates the effects with slow and
  fast dependence on time. The resulting closure approximates the
  unknown quantity by the fluctuations of the single-particle velocity
  divergence around its own time average. The latter, in turn, is
  connected to the pressure variable via the Green--Kubo formula. This
  is described in Section~\ref{sec:trJ_closure}.
\item The novel closure introduces a combination of linear damping (at
  small scales) and viscous diffusion (at large scales) into the
  pressure equation, which stabilizes the pressure solution. In
  Section~\ref{sec:properties}, we study basic properties of the
  damped pressure equation. The key findings are as follows:
  \begin{enumerate}[label=\alph*)]
  \item We examine the wave structure of linearized solutions, and
    find that, while the acoustic waves are no longer present at
    physically relevant scales, the novel ``density wave'' solutions
    emerge due to the presence of the van der Waals effect. Throughout
    the troposphere, the phase speed of such waves varies roughly
    between 5--15 m/s, which anecdotally matches the observed speeds
    of propagation of the atmospheric planetary waves, such as the
    equatorial Rossby and Kelvin waves, as well as the Madden--Julian
    oscillation.
  \item At low Mach numbers, the novel dissipative effect can be used
    to render the pressure equation diagnostic via the averaging
    formalism, while the density and velocity variables remain
    prognostic. This effectively confers the combination of linear
    damping (at large scales) and viscous diffusion (at small scales)
    onto the velocity divergence.  The corresponding diffusion
    coefficient is known as the {\em second viscosity} \cite[Section
      81]{LanLif}. The ratio of the second viscosity and the usual
    shear viscosity is inversely proportional to the packing fraction,
    which, at normal conditions, is $\sim 6.5\cdot 10^{-4}$
    \cite{Abr26}. As a result, at normal conditions, the second
    viscosity is about five hundred times greater than the shear
    viscosity, which is corroborated by some measurements
    \cite{Cramer2012}.
    \item It is known from observations that turbulent dynamics emerge
      spontaneously when the Reynolds number is of the order $\sim
      2\cdot 10^3$ \cite{Rey83,Rey,Let,Men}. In our model, the linear
      instability, which creates turbulent dynamics, is governed by
      the second viscosity, and the critical value of the Reynolds
      number corresponds to the ratio of the second and shear
      viscosities --- that is, the reciprocal of the packing fraction.
      This matches observations to an order of magnitude.
  \end{enumerate}
\end{enumerate}
We find it remarkable that neither the second viscosity, nor the
Reynolds criterion appear to be linked directly to the momentum
diffusivity via the Newton law of viscosity, being instead averaged
effects of the pressure dynamics at low Mach numbers.

\section{Preliminaries: from the Newton equations to fluid mechanics}

Adopting the standard approach of kinetic theory, we start with a
system of $K$ particles in a domain of volume $V$, with their
coordinates and velocities at a time $t$ denoted via $\BV x_i(t)$ and
$\BV v_i(t)$, respectively, with $i=1,\ldots,K$. The particles
interact with each other via a potential $\phi(r)$, with $r$ being the
distance between the interacting particles; for simplicity of
calculations, particles are presumed to lack rotational or vibrational
degrees of freedom.  Such a motion is described by the following
system of Newton's equations:
\begin{equation}
\label{eq:ODE_sys}
\deriv{\BV x_i}t=\BV v_i,\qquad\deriv{\BV v_i}t=-\parderiv{}{\BV x_i}
\sum_{\myatop{j=1}{j\neq i}}^K\phi(\|\BV x_i-\BV x_j\|).
\end{equation}
We assume that the potential has a finite range $\sigma$, that is,
$\phi(r)=0$ for $r>\sigma$. Additionally, $\phi(r)>0$ as $r\to 0$,
that is, the potential is overall repelling.

The average momentum $\BV u_0$ of the system in \eqref{eq:ODE_sys} is
given via
\begin{equation}
\BV u_0=\frac 1K\sum_{i=1}^K\BV v_i.
\end{equation}
It is preserved in time irrespectively of what $\phi(r)$ is:
\begin{equation}
\deriv{\BV u_0}t=\frac 1K\sum_{i=1}^K\deriv{\BV v_i}t=\frac 1K
\sum_{i=1}^{K-1}\sum_{j=i+1}^K\bigg(\parderiv{}{\BV x_i}+\parderiv{
}{\BV x_j}\bigg)\phi(\|\BV x_i-\BV x_j\|)=\BV 0.
\end{equation}
To show the total energy conservation, it is convenient to denote
\begin{equation}
\BV X=(\BV x_1,\ldots,\BV x_K),\qquad\BV V=(\BV v_1,\ldots,\BV v_K),
\qquad\Phi(\BV X)=\sum_{i=1}^{K-1}\sum_{j=i+1}^K\phi(\|\BV x_i-\BV x_j
\|).
\end{equation}
Then, the system of equations in \eqref{eq:ODE_sys} can be expressed
via
\begin{equation}
\label{eq:ODE}
\deriv{\BV X}t=\BV V,\qquad\deriv{\BV V}t=-\parderiv\Phi{\BV X}.
\end{equation}
Introducing the total energy
\begin{equation}
\label{eq:E}
E(\BV X,\BV V)=\frac 12\|\BV V\|^2+\Phi(\BV X),
\end{equation}
we can calculate
\begin{equation}
\deriv{}tE(\BV X(t),\BV V(t))=\parderiv E{\BV X}\cdot\deriv{\BV X}t
+\parderiv E{\BV V} \cdot\deriv{\BV V}t=\parderiv\Phi{\BV X}\cdot\BV
V-\BV V\cdot \parderiv\Phi{\BV X}=0.
\end{equation}
Remarkably, $E$ is the Hamiltonian of \eqref{eq:ODE}, with $\BV X$ and
$\BV V$ being the canonical variables:
\begin{equation}
\deriv{\BV X}t=\parderiv E{\BV V},\qquad\deriv{\BV V}t=-\parderiv
E{\BV X}.
\end{equation}

\subsection{The density of states and the Liouville equation}

Here, we introduce the density of states $F(t,\BV X,\BV V)$ of the
system in \eqref{eq:ODE}. This density satisfies the Liouville
equation
\begin{equation}
\label{eq:Liouville}
\parderiv Ft+\BV V\cdot\parderiv F{\BV X}=\parderiv\Phi{\BV X}\cdot
\parderiv F{\BV V}.
\end{equation}
The derivation of \eqref{eq:Liouville} via the infinitesimal generator
of \eqref{eq:ODE} is given, for instance, in \cite{Abr24}. For
convenience, we assume that the coordinate domain has no boundary
effects, such that the integration by parts does not introduce
boundary terms. Now, we look at some properties of solutions of
\eqref{eq:Liouville}.

Let $\psi:\RR\to\RR$ be a differentiable function. Then, the integral
of $\psi(F)$ is preserved in time:
\begin{equation}
\parderiv{}t\int_{\RR^{3K}}\int_{V^K}\psi(F)\dif\BV X\dif\BV V=\int_{
  \RR^{3K}}\int_{V^K}\left[\parderiv{}{\BV V}\cdot\left(\psi(F)
  \parderiv\Phi{\BV X}\right)-\parderiv{}{\BV X}\cdot\big(\psi(F)\BV V
  \big)\right]\dif\BV X\dif\BV V=0.
\end{equation}
In particular, any $L^p$-norm of $F$ is preserved in time as a
consequence. Moreover, the $L^\infty$-norm of $F$ is also preserved in
time; for that, observe that $F$ remains constant on a characteristic
of \eqref{eq:Liouville}:
\begin{equation}
\deriv{}tF(t,\BV X(t),\BV V(t))=\parderiv Ft+\parderiv F{\BV X}\cdot
\deriv{\BV X}t+\parderiv F{\BV V}\cdot\deriv{\BV V}t=\parderiv Ft+
\parderiv F{\BV X}\cdot\BV V-\parderiv F{\BV V}\cdot\parderiv\Phi{\BV
  X}=0.
\end{equation}
Additionally, given a steady state $F_0$ of \eqref{eq:Liouville}, a
generic solution $F$ of \eqref{eq:Liouville} preserves the R\'enyi
divergence family $D_\alpha(F,F_0) $\cite{Ren} (see \cite{Abr22} for
details), including the Kullback--Leibler divergence \cite{KulLei}.

There are multiple steady states $F_0$ of \eqref{eq:Liouville}. One of
them, known as the Gibbs canonical equilibrium state, is given via
\begin{equation}
\label{eq:Gibbs}
F_G(\BV X,\BV V)=\frac{e^{-\Phi(\BV X)/\theta}}{Z_K}\frac{\prod_{i=1}^
  K e^{ -\|\BV v_i-\BV u_0\|^2/2\theta}}{(2\pi\theta)^{3K/2}},\qquad
Z_K =\int_{V^K} e^{-\Phi(\BV X)/\theta}\dif\BV X,
\end{equation}
where the parameter $\theta$ is the equilibrium kinetic temperature of
the system of particles,
\begin{equation}
\label{eq:eqT}
\theta=\frac 13\int_{\RR^{3K}}\int_{V^K}\|\BV v_i-\BV u_0\|^2 F_G(\BV
X,\BV V)\dif\BV X\dif\BV V,\quad\forall i,\; 1\leq i\leq K.
\end{equation}
The Gibbs equilibrium state $F_G$ maximizes the Shannon entropy
\cite{Shan} under the prescribed average momentum and total energy
constraints, and is thus regarded as the most ``statistically common''
equilibrium state encountered in nature. The single-particle density
$f_G$ and the joint two-particle density $f_G^{(2)}$ are defined,
respectively, via
\begin{equation}
f_G(\BV x_i,\BV v_i)=\int_{\RR^{3(K-1)}}\int_{V^{K-1}}F_G\prod_{
  \myatop{j=1}{j\neq i}}^K\dif\BV x_j\dif\BV v_j,\quad f_G^{(2)}(\BV
x_i,\BV v_i,\BV x_j,\BV v_j)=\int_{\RR^{3(K-2)}}\int_{V^{K-2}}F_G
\prod_{\myatop{k=1}{k\neq i,j}}^K\dif\BV x_k\dif\BV v_k.
\end{equation}
The explicit formulas for $f_G$ and $f_G^{(2)}$ are given,
respectively, via
\begin{subequations}
\begin{equation}
\label{eq:fG}
f_G(\BV v)=\frac{e^{-\|\BV v-\BV u_0\|^2/2\theta}}{(2\pi\theta)^{3/2}V
},
\end{equation}
\begin{equation}
\label{eq:fG2}
f_G^{(2)}(\BV x,\BV v,\BV y,\BV w)=e^{-\frac{\phi(\|\BV x-\BV y\|)}
  \theta}Y_K(\|\BV x-\BV y\|)f_G(\BV v)f_G(\BV w),
\end{equation}
\end{subequations}
where $\BV x$, $\BV v$, $\BV y$ and $\BV w$ are the coordinates and
velocities of the two particles. Above, $Y_K(r)$ is the pair cavity
distribution function for $K$ particles \cite{Abr24,Bou86}, given via
\begin{equation}
\label{eq:Y}
Y_K(\|\BV x-\BV y\|)=\frac{V^2}{Z_K}\int_{V^{K-2}}\prod_{i=3}^K e^{-(
  \phi(\|\BV x-\BV x_i\|)+\phi(\|\BV y-\BV x_i\|))/\theta}\left(
\prod_{j=i+1}^K e^{-\phi(\|\BV x_i-\BV x_j\|)/\theta}\right)\dif\BV
x_i.
\end{equation}

\subsection{Transport equation for the distribution function of a single
particle}

Here we follow the standard BBGKY formalism \cite{Bog,BorGre,Kir,Yvo}
to obtain the approximate transport equation for the density of a
single particle. Let us integrate the Liouville equation in
\eqref{eq:Liouville} over all particles but one:
\begin{equation}
\left(\parderiv{}t+\BV v\cdot\parderiv{}{\BV x}\right)f_i(t,\BV x,\BV
v)=-\parderiv{}{\BV v}\cdot\sum_{\myatop{j=1}{j\neq i}}^K\int_{\RR^3}
\int_{B(\sigma)}\parderiv{\phi(\|\BV z\|)}{\BV z}f_{ij}^{(2)}(t,\BV
x,\BV v,\BV x+\BV z,\BV w)\dif\BV z\dif\BV w,
\end{equation}
where $\BV z=\BV y-\BV x$ is a dummy variable of spatial integration,
and $B(\sigma)$ is a ball of radius $\sigma$. The single-particle
distribution $f_i(t,\BV x,\BV v)$ and the joint two-particle
distribution $f_{ij}^{(2)}(t,\BV x,\BV v,\BV y,\BV w)$ are given via
\begin{equation}
f_i(t,\BV x_i,\BV v_i)=\int_{\RR^{3(K-1)}}\int_{V^{K-1}} F\prod_{
  \myatop{j=1}{j\neq i}}^K\dif\BV x_j\dif\BV v_j,\quad f_{ij}^{(2)}
(t,\BV x_i,\BV v_i,\BV x_j,\BV v_j)=\int_{\RR^{3(K-2)}}\int_{V^{K-2}}
F\prod_{\myatop{k=1}{k\neq i,j}}^K\dif\BV x_k\dif\BV v_k.
\end{equation}
Clearly,
\begin{equation}
f_i(t,\BV x,\BV v)=\int_{\RR^3}\int_V f_{ij}^{(2)}(t,\BV x,\BV v,\BV
y,\BV w)\dif\BV y\dif\BV w=\int_{\RR^3}\int_V f_{ji}^{(2)}(t,\BV y,\BV
w,\BV x,\BV v)\dif\BV y\dif\BV w,
\end{equation}
for all distinct $i$ and $j$. Since it is impossible to track
statistical properties of individual particles, we set
\begin{equation}
\label{eq:f_f2_generic}
f_i(t,\BV x,\BV v)=f(t,\BV x,\BV v),\qquad f_{ij}^{(2)}(t,\BV x,\BV
v,\BV y,\BV w)=f^{(2)}(t,\BV x,\BV v,\BV y,\BV w),
\end{equation}
where $f$ and $f^{(2)}$ are the distributions of a ``generic''
particle and a pair of particles, respectively. The resulting,
approximate, transport equation for $f$ is given via
\begin{equation}
\label{eq:transport_equation}
\left(\parderiv{}t+\BV v\cdot\parderiv{}{\BV x}\right)f(t,\BV x,\BV v)
=-(K-1)\parderiv{}{\BV v}\cdot\int_{\RR^3}\int_{B(\sigma)}\parderiv{
  \phi(\|\BV z\|)}{\BV z}f^{(2)}(t,\BV x,\BV v,\BV x+\BV z,\BV w)\dif
\BV z\dif\BV w.
\end{equation}
The next step is to achieve a {\em closure}, that is, to express
$f^{(2)}$ via $f$. Generally, this is achieved by defining the {\em
  pair correlation function} \cite{vBeiErn,Bou86} $g(\BV x,\BV v,\BV
y,\BV w)$ via
\begin{equation}
\label{eq:f2_structure}
f^{(2)}(\BV x,\BV v,\BV y,\BV w)=f(\BV x,\BV v)f(\BV y,\BV w)
g(\BV x,\BV v,\BV y,\BV w).
\end{equation}
The transport equation for $f$ in \eqref{eq:transport_equation} thus
becomes
\begin{equation}
\parderiv ft+\BV v\cdot\parderiv f{\BV x}=-(K-1)\parderiv{}{\BV v}
\cdot\int_{\RR^3}\int_{B(\sigma)}\parderiv{\phi(\|\BV z\|)}{\BV z}g(
\BV x,\BV v,\BV x+\BV z,\BV v)f(\BV x,\BV v)f(\BV x+\BV z,\BV w)
\dif\BV z\dif\BV w.
\end{equation}
The next step is to introduce the mass of a particle, $m$, and rescale
$f\to Kmf$ so that $f$ becomes the mass density. Assuming that $K$ is
large enough so that $(K-1)/K\to 1$, we arrive at the following
transport equation for the single-particle distribution function $f$:
\begin{subequations}
\label{eq:f_transport_collision}
\begin{equation}
\label{eq:f_transport}
\parderiv ft+\BV v\cdot\parderiv f{\BV x}=\parderiv{}{\BV v}\cdot
\coll[f],
\end{equation}
\begin{equation}
\label{eq:f_collision}
\coll[f]=-\frac 1mf(\BV x,\BV v)\int_{\RR^3}\int_{B(\sigma)}\parderiv{
  \phi(\|\BV z\|)}{\BV z}g(\BV x,\BV v,\BV x+\BV z,\BV w)f(\BV x+\BV
z,\BV w)\dif\BV z\dif\BV w.
\end{equation}
\end{subequations}
This is a Vlasov-type equation \cite{Vla}, because the collision
integral is time-reversible. Due to the time reversibility, this
collision integral does not damp the nonequilibrium higher-order
velocity moments such as the stress and heat flux, and thereby cannot
describe thermodynamically irreversible effects such as viscosity and
heat conduction.  However, here we resort to the Vlasov collision
integral due to the convenience of calculations to follow; the
Boltzmann collision integral will be studied in the future
work. Subsequently, the higher-order nonequilibrium effects such as
the stress and heat flux will be computed from empirically observed
constitutive relations, such as the Newton law of viscosity, and the
Fourier law of heat conduction.

\subsection{Transport equations for the mass density, momentum and
pressure}

For a function $\psi(\BV v)$, let us define the corresponding velocity
moments of $f$ and $\coll[f]$ via
\begin{equation}
\langle\psi(\BV v)\rangle_f(t,\BV x)=\int_{\RR^3}\psi(\BV v)f(t,\BV x,
\BV v)\dif\BV v,\quad\langle\psi(\BV v)\rangle_{\coll[f]}(t,\BV x)=
\int_{\RR^3}\psi(\BV v)\coll[f](t,\BV x,\BV v) \dif\BV v.
\end{equation}
The transport equation for $\langle\psi\rangle_f$ is computed by
integrating \eqref{eq:f_transport} against $\psi(\BV v)$,
\begin{equation}
\label{eq:m_transport}
\parderiv{}t\langle\psi\rangle_f+\nabla\cdot\langle\psi\BV v\rangle_f
=-\left\langle\parderiv\psi{\BV v}\cdot\right\rangle_{\coll[f]},
\end{equation}
where the $\BV v$-derivative in the right-hand side was integrated by
parts. The dot in the collision moment denotes the scalar
multiplication of $\partial\psi/\partial\BV v$ by $\coll[f]$, as the
latter is a vector.  The moment equation above is not closed with
respect to $\langle\psi\rangle_f$, as the advection term contains the
higher-order velocity moment $\langle\psi\BV v\rangle_f$. Thus, the
equations for velocity moments of different orders are chain-linked to
each other, creating a hierarchy.

The low-order velocity moments of $f$ are the density $\rho$,
velocity $\BV u$, and pressure $p$:
\begin{equation}
\label{eq:rho_u_p}
\rho=\langle 1\rangle_f,\qquad\BV u=\frac 1\rho\langle\BV v\rangle_f,
\qquad p=\frac 13\langle\|\BV v-\BV u\|^2\rangle_f.
\end{equation}
The kinetic temperature is $\theta=p/\rho$. From
\eqref{eq:m_transport}, we obtain the equations for $\rho$, $\BV u$
and $p$:
\begin{subequations}
\label{eq:rho_u_p_coll}
\begin{equation}
\label{eq:rho_u_coll}
\Dderiv\rho t+\rho\nabla\cdot\BV u=0,\qquad\rho\Dderiv{\BV u}t+\nabla
p+\nabla\cdot\BV\Sigma=-\langle\BM I\cdot\rangle_{\coll[f]},
\end{equation}
\begin{equation}
\Dderiv pt+\frac 53p \nabla\cdot\BV u+\frac 23(\BV\Sigma:\nabla\BV
u+\nabla\cdot\BV q)=-\frac 23\left\langle(\BV v-\BV u)\cdot\right
\rangle_{\coll[f]},
\end{equation}
\end{subequations}
where $\BV\Sigma$ and $\BV q$ are the shear stress and heat flux,
respectively:
\begin{equation}
\label{eq:stress_heat_flux}
\BV\Sigma=\langle(\BV v-\BV u)^2\rangle_f-p\BM I,\qquad \BV q=\frac
12\langle\|\BV v-\BV u\|^2(\BV v-\BV u)\rangle_f.
\end{equation}
The derivation of the transport equations above in
\eqref{eq:rho_u_p_coll} is given in Appendix
\ref{sec:moment_derivation}.  We note that the usual compressible
Euler equations \eqref{eq:Euler} for a monatomic gas ($\gamma=5/3$)
are obtained from \eqref{eq:rho_u_p_coll} by setting $\BV\Sigma$, $\BV
q$ and both collision integrals to zero. In what follows, we set the
stress $\BV\Sigma$ and the heat flux $\BV q$ in
\eqref{eq:stress_heat_flux} to what is observed in nature. Namely, at
normal conditions, $\BV\Sigma$ and $\BV q$ obey the Newton law of
viscosity, and the Fourier law of heat conduction combined with the
radiative cooling, respectively:
\begin{equation}
\label{eq:Newton_Fourier}
\BV\Sigma=-\mu\left(\nabla\BV u+\nabla\BV u^T-\frac 23(\nabla\cdot\BV
u)\BM I\right),\qquad\nabla\cdot\BV q=-\nabla\cdot(\kappa\nabla T)
+4\alpha\sigma_{SB}(T^4-T_0^4).
\end{equation}
Above, $\mu$ is the dynamic viscosity, $\kappa$ is the heat
conductivity, $\sigma_{SB}$ is the Stefan--Boltzmann constant,
$\alpha$ is the electromagnetic absorption coefficient, and $T_0$ is
the background temperature. The radiative cooling is computed in the
optically thin limit of the differential approximation to the equation
of transfer (see, for example, p.~106 of \cite{SieHow}, eq.~(1-113)),
under the assumption that the gas is largely transparent. For the sake
of simplicity, we treat $\mu$, $\kappa$ and $\alpha$ as constants
throughout the rest of the work.

\section{A correction in the pair correlation function}
\label{sec:closure}

Conventionally, the pair correlation function $g$ in
\eqref{eq:f2_structure} is represented by a radial distribution
function \cite{vBeiErn,Abr17,Abr22,Abr24,Bou86}, which accounts for
collisional interactions and depends solely on the distance between
the colliding particles. In particular, setting $g$ to the pair
correlation function for the Gibbs joint state $f_G^{(2)}$ in
\eqref{eq:fG2},
\begin{equation}
\label{eq:conventional_collision}
g_0=\exp\left(-\frac{\phi(\|\BV x-\BV y\|)}\theta\right)Y(\|\BV x-\BV
y \|),
\end{equation}
where $Y$ refers to the pair cavity distribution function for
infinitely many particles, leads to a Vlasov-type equation \cite{Vla},
and, subsequently, to the usual hierarchy of the velocity moment
equations with the standard transport equations for the density,
momentum and pressure \cite{Abr22}. Moreover, an assumption that $f$
is a Gaussian distribution \eqref{eq:fG} with a prescribed density
$\rho$, velocity $\BV u$ and pressure $p$, leads directly to the
compressible Euler equations \eqref{eq:Euler}. For the form of $g$ in
\eqref{eq:conventional_collision}, there are no effects which
correspond to the stabilized pressure behavior as observed in nature
at low Mach numbers.

In what follows, we propose a correction to the standard pair
correlation function in~\eqref{eq:conventional_collision}, which
accounts for variations in average velocity of particles at
neighboring locations. As a consequence of that, the corrected pair
correlation function can describe effects pertaining to the
compression or expansion rate of the gas, which are missing from the
standard pair correlation function in
\eqref{eq:conventional_collision}.

\subsection{Bridging the mean interparticle distance and the gas
  compression/expansion rate}
\label{sec:mean_distance}

Consider a particle at the location $\BV x$, surrounded by $K-1$
particles at locations $\BV x_i$, $2\leq i\leq K$.  The mean distance
$\langle D\rangle$ between the particle at $\BV x$ and all other
particles is
\begin{equation}
\label{eq:D_x}
\langle D\rangle=\frac 1{K-1}\sum_{i=2}^K\|\BV x_i-\BV x\|.
\end{equation}
Next, we assume that the particles are distributed spatially uniformly
in a ball $B(R)$ of radius $R$ and volume $V$, centered at $\BV x$.
Then, as $K\to\infty$, the expectation $\EE\langle D\rangle$ is
\begin{equation}
\EE\langle D\rangle=\frac 1V\int_{B(R)}\|\BV z\|\dif\BV z=\frac{4\pi}V
\int_0^Rr^3\dif r=\frac{\pi R^4}V=\left(\frac 34\right)^{4/3}
\frac{V^{1/3}}{\pi^{1/3}},
\end{equation}
where $\BV z=\BV x_i-\BV x$. In reality, particles are unlikely to
approach each other closer than the effective range $\sigma$ of the
potential $\phi$, however, we assume that the gas is dilute (that is,
$\|\BV x_i-\BV x\|\gg\sigma$ on average), and thus the potential
interactions can be neglected.

Next, we recall that each particle has the mass $m$. Then, the mass
density $\rho$ is
\begin{equation}
\rho=\frac{Km}V.
\end{equation}
Expressing $V$ via $\rho$, we obtain
\begin{equation}
\label{eq:D_rho}
\EE\langle D\rangle=\left(\frac 34\right)^{4/3}\left(\frac{Km}{\pi\rho
}\right)^{1/3}.
\end{equation}
Next, we examine the rate of change of $\langle D\rangle$ due to the
particle movement. From \eqref{eq:D_rho}, we have
\begin{equation}
\label{eq:D_divu}
\deriv{}t\EE\langle D\rangle=\left(\frac 34\right)^{4/3}\left(\frac{Km
}{\pi\rho}\right)^{1/3}\left(-\frac 13\right)\frac 1\rho\Dderiv\rho t=
\frac 14\left(\frac{3V}{4\pi}\right)^{1/3}\nabla\cdot\BV u.
\end{equation}
The last equality follows from the mass conservation equation in
\eqref{eq:rho_u_coll}. Also, it is tacitly assumed that
$\nabla\cdot\BV u$ is constant in $B(R)$.  From \eqref{eq:D_divu}, it
follows that the average distance between particles increases if the
gas expands ($\nabla\cdot\BV u>0$), and vice versa.

Now, we model the same behavior using the closure for $f^{(2)}$ via
\eqref{eq:conventional_collision}. The time derivative of $\langle
D\rangle$, expressed via the infinitesimal generator $\cL$ of
\eqref{eq:ODE_sys}, is
\begin{equation}
\deriv{\langle D\rangle}t=\frac 1{K-1}\sum_{i=2}^K\deriv{\|\BV x_i-\BV
  x\|}t=\frac 1{K-1}\sum_{i=2}^K\frac{\BV x_i-\BV x}{\|\BV x_i-\BV
  x\|}\cdot(\BV v_i-\BV v)\equiv\cL\langle D\rangle.
\end{equation}
The particles are distributed according to the probability density $F$
from the Liouville equation \eqref{eq:Liouville}, where we denote $\BV
x=\BV x_1$, and $\BV v=\BV v_1$. Note that $\langle D\rangle$ is
conditional on $\BV x$, but not on $\BV v$ (that is, we assume that
the first particle is at $\BV x$, but its velocity is unspecified).
Therefore, the time derivative of the expectation of $\langle
D\rangle$ is
\begin{equation}
\deriv{}t\EE\langle D\rangle=\frac{\int_{\RR^{3K}}\int_{B(R)^{K-1}}\cL
  \langle D\rangle F\dif\BV x_2\ldots\dif\BV x_K\dif\BV V}{\int_{
    \RR^{3K}}\int_{B(R)^{K-1}} F\dif\BV x_2\ldots\dif\BV x_K\dif\BV
  V},
\end{equation}
where the denominator is $V^{-1}$ (as the spatial distribution of an
unspecified particle is taken to be uniform), and
\begin{multline}
\int_{\RR^{3K}}\int_{B(R)^{K-1}}\cL\langle D\rangle F\dif\BV x_2\ldots
\dif\BV x_K\dif\BV V\\=\frac 1{K-1}\sum_{i=2}^K\int_{\RR^6}\int_{B(R)}
\frac{\BV z}{\|\BV z\|}\cdot(\BV w-\BV v)f^{(2)}_{1i}(\BV x,\BV v,\BV
x+\BV z,\BV w)\dif\BV z\dif\BV v\dif\BV w.
\end{multline}
We now assume that all pairs are statistically equivalent, which
implies \eqref{eq:f_f2_generic}, and yields
\begin{equation}
\label{eq:F_divu}
\deriv{}t\EE\langle D\rangle=V\int_{\RR^6}\int_{B(R)}\frac{\BV z}{
  \|\BV z\| }\cdot(\BV w-\BV v)f^{(2)}(\BV x,\BV v,\BV x+\BV z,\BV
w)\dif\BV z\dif \BV v\dif\BV w.
\end{equation}
If $f^{(2)}$ above is the pair marginal of the solution $F$ of
\eqref{eq:Liouville}, then \eqref{eq:F_divu} equals \eqref{eq:D_divu}:
\begin{equation}
\label{eq:DF_divu}
V\int_{\RR^6}\int_{B(R)}\frac{\BV z}{\|\BV z\|}\cdot(\BV w-\BV v)
f^{(2)}(\BV x,\BV v,\BV x+\BV z,\BV w)\dif\BV z\dif\BV v\dif\BV w=
\frac 14\left(\frac{3V}{4\pi}\right)^{1/3}\nabla\cdot\BV u.
\end{equation}
However, if $f^{(2)}$ is approximated by a closure, then
\eqref{eq:DF_divu} may not necessarily hold.

The conventional closure for $f^{(2)}$ in kinetic theory consists of
\eqref{eq:f2_structure} paired with \eqref{eq:conventional_collision}.
Above in Section~\ref{sec:mean_distance}, we assumed that the
particles are distributed spatially uniformly, which means that $f(\BV
x,\BV v)=f(\BV v)$. However, in this case the integral in the
left-hand side of~\eqref{eq:DF_divu} vanishes upon integration over
$\dif\BV z$ alone:
\begin{multline}
V\int_{\RR^6}\int_{B(R)}\frac{\BV z}{\|\BV z\|}\cdot(\BV w-\BV v)
f^{(2)}(\BV x,\BV v,\BV x+\BV z,\BV w)\dif\BV z\dif\BV v\dif\BV w\\=V
\int_{\RR^6}\int_{B(R)}\frac{\BV z}{\|\BV z\|}\cdot(\BV w-\BV v)e^{
  -\frac{\phi(\|\BV z\|)}\theta}Y(\|\BV z\|)f(\BV v)f(\BV w)\dif\BV z
\dif\BV v\dif\BV w\\=V\int_0^R e^{-\frac{\phi(r)}\theta}Y(r)r^2\dif r
\int_{\mathbb S_1}\BV n\dif\BV n\cdot\int_{\RR^6}(\BV w-\BV v)f(\BV v)
f(\BV w)\dif\BV v\dif\BV w=0.
\end{multline}
Above, we switched to the spherical coordinate system $\BV z=r\BV n$,
$\dif\BV z=r^2\dif r\dif\BV n$, where $\BV n$ is a vector on the unit
sphere $\mathbb S_1$. The integration of $\BV n\dif\BV n$ over the
sphere cancels out.

\subsection{The proposed correction to the pair correlation function}
\label{sec:correction}

The structure of \eqref{eq:DF_divu} suggests that the pair correlation
function $g$ depends on $(\BV v-\BV w)$, in addition to $\BV x$ and
$\BV y$, however, it is unclear which form such a correction would
take. In order to guess this form, we examine the correction to the
Gibbs state \eqref{eq:fG2} of a two-particle system under the
assumption that the velocity $\BV u_0$ is variable (but the
temperature $\theta$ is constant).

For a generic pair of two particles, let $\BV U(\BV x,\BV
y):\RR^6\to\RR^3$ denote the average velocity of the particle at $\BV
x$, given that the other particle is at $\BV y$. Note that if the two
particles are independent, then the dependence of $\BV U$ on its
second argument vanishes. The Gaussian state, which corresponds to
such two particles, is given via
\begin{equation}
f^{(2)}_{\BV U}=\frac{Y(\|\BV x-\BV y\|)}{(2\pi\theta)^3V^2}\exp\left(
-\frac{\|\BV v-\BV U(\BV x,\BV y)\|^2}{2\theta}-\frac{\|\BV w-\BV
  U(\BV y,\BV x)\|^2 }{2 \theta}-\frac{\phi(\|\BV x-\BV
  y\|)}\theta\right) .
\end{equation}
Indeed, observe that the average velocities of each particle are given
via
\begin{equation}
\langle\BV v\rangle=\frac{\int_{\RR^6}\BV vf^{(2)}_{\BV U}\dif\BV v
  \dif\BV w}{\int_{\RR^6}f^{(2)}_{\BV U}\dif\BV v\dif\BV w}=\BV U(\BV
x,\BV y),\qquad\langle\BV w\rangle=\frac{\int_{\RR^6}\BV wf^{(2)}_{\BV
    U}\dif\BV v\dif\BV w}{\int_{\RR^6}f^{(2)}_{\BV U}\dif\BV v\dif\BV
  w}=\BV U(\BV y,\BV x),
\end{equation}
since the ``potential wells'' due to $\phi$ are present both in the
numerator and denominator and therefore cancel out. Next, we denote
\begin{equation}
\label{eq:Deltau}
\delta\BV U(\BV x,\BV y)=\BV U(\BV y,\BV x)-\BV U(\BV x,\BV y),
\end{equation}
and, using $\BV U$ as a shorthand for $\BV U(\BV x,\BV y)$, rewrite
the expression $\|\BV w-\BV U(\BV y,\BV x)\|^2$ as
\begin{multline}
\|\BV w-\BV U(\BV y,\BV x)\|^2=\|\BV w-\BV U+\BV U-\BV U(\BV y,\BV
x)\|^2=\|\BV w-\BV U\|^2- 2(\BV w-\BV U)\cdot\delta\BV U\\+\|\delta\BV
U\|^2=\|\BV w-\BV U\|^2 +(\BV v-\BV w)\cdot\delta\BV U-(\BV w+\BV
v-2\BV U)\cdot\delta\BV U +\|\delta\BV U\|^2,
\end{multline}
where in the last identity we added and subtracted $\BV v$ in the
expression $2(\BV w-\BV U)\cdot\delta\BV U$. With this, $f^{(2)}_{\BV
  U}$ can be expressed via
\begin{subequations}
\begin{equation}
f^{(2)}_{\BV U}=f^{(2)}_{\BV U,G}\exp\left(\frac A{2\theta}+\frac B{
  2\theta}-\frac{\|\delta\BV U\|^2}{2\theta}\right)\\=f^{(2)}_{\BV U,G
}\left(1+\frac A{2\theta}+\frac B{2\theta}\right)+O(\|\delta\BV U\|^2),
\end{equation}
\begin{equation}
A=(\BV w-\BV v)\cdot\delta\BV U,\qquad B=(\BV w+\BV v-2\BV
u)\cdot\delta\BV U,
\end{equation}
\end{subequations}
where $f^{(2)}_{\BV U,G}$ is the two-particle Gibbs equilibrium state
\eqref{eq:fG2} with the average velocity~$\BV U(\BV x,\BV y)$ for both
particles. Henceforth we assume that the variations in $\BV U(\BV
x,\BV y)$ are small enough so that the quadratic term $O(\|\delta\BV
U\|^2)$ can be ignored in the computations of the remainder of this
section. Next, we substitute $f^{(2)}_{\BV U}$ above into the integral
in \eqref{eq:DF_divu}.  Integrating the two corrections $A$ and $B$ to
$f^{(2)}_{ \BV U,G}$ separately, for the integral over the velocities
alone in the correction $B$ we obtain
\begin{multline}
\int_{\RR^6}(\BV w-\BV v)(\BV w +\BV v-2\BV U)^Tf^{(2)}_{\BV U,G}\dif
\BV v\dif\BV w\\=\int_{\RR^6}(\BV w-\BV U-\BV v+\BV U)(\BV w-\BV U+\BV
v-\BV U)^Tf^{(2)}_{\BV U,G}\dif\BV v\dif\BV w\\\!=\int_{\RR^6}\left[(\BV
  w-\BV U)^2-(\BV v-\BV U)^2+(\BV w-\BV U)(\BV v-\BV U)^T-(\BV v-\BV U
  )(\BV w-\BV U)^T\right]f^{(2)}_{\BV U,G}\dif\BV v\dif\BV w=\BM 0,
\end{multline}
and thus the correction $B$ has no effect in \eqref{eq:DF_divu}. For
the correction $A$ we obtain
\begin{equation}
V\int_{\RR^6}\int_V\frac{\BV z}{\|\BV z\|}\cdot(\BV w-\BV v)\frac A{2
  \theta}f^{(2)}_{\BV U,G}\dif\BV z\dif\BV v\dif\BV w=\frac V{2\theta}
\int_{\RR^6}\int_V\frac{\BV z^T}{\|\BV z \|}(\BV w-\BV v)^2\delta\BV U
f^{(2)}_{\BV U,G}\dif\BV z\dif\BV v \dif\BV w.
\end{equation}
The integral over the velocities separately is
\begin{multline}
\int_{\RR^6}(\BV w-\BV v)^2f^{(2)}_{\BV U,G}\dif\BV v\dif\BV w=\int_{
  \RR^6}\left[(\BV w-\BV U-\BV v+\BV U)^2\right]f^{(2)}_{\BV U,G}\dif
\BV v\dif \BV w\\=\int_{\RR^6}\Big[(\BV w-\BV U)^2+(\BV v-\BV U)^2+
  (\BV w-\BV U) (\BV v-\BV U)^T+(\BV v-\BV U)(\BV w-\BV U)^T\Big]
f^{(2)}_{\BV U,G} \dif\BV v\dif\BV w\\=\frac{2\theta}{V^2}
e^{-\frac{\phi(\|\BV x-\BV y\|)} \theta}Y(\|\BV x-\BV y\|)\BM I,
\end{multline}
and the dependence on $\BV U(\BV x,\BV y)$ vanishes, although the
velocity difference $\delta\BV U(\BV x,\BV y)$ still remains in the
integral.  Subsequently,
\begin{multline}
\label{eq:f2_corr}
\frac V{2\theta}\int_{\RR^6}\int_{B(R)}\frac{\BV z^T}{\|\BV z\|}(\BV w
-\BV v)^2\delta\BV U f^{(2)}_{\BV U,G}\dif\BV z\dif\BV v\dif\BV w\\=
\frac 1V\int_{B(R)}\frac{\BV z\cdot\delta\BV U}{\|\BV z\|}e^{-\frac{
    \phi(\|\BV z\|)}\theta}Y(\|\BV z\|)\dif\BV z=\frac 1V\int_{B(R)}
\frac{\BV z\cdot\delta\BV U}{\|\BV z\|}\dif\BV z=\frac 1V\int_0^R
\left(\int_{\mathbb S_1}\delta\BV U\cdot\BV n r^2\dif\BV n\right)\dif
r,
\end{multline}
where in the second line we discarded the potential well (because, for
a dilute gas, it is too narrow to affect the integral), and expressed
the integral in the spherical coordinates.

To evaluate the integral above, we apply the Gauss theorem to the
surface integral:
\begin{equation}
\int_{\mathbb S_1}\delta\BV U\cdot\BV nr^2\dif\BV n=\int_{\mathbb S(r)
}\delta\BV U\cdot\BV n\dif A=\int_{B(r)}\nabla_{\BV z}\cdot\delta\BV
U(\BV x,\BV x+\BV z)\dif\BV z,
\end{equation}
where the area integral over the sphere $\mathbb S(r)$ of radius $r$
is transformed into the volume integral over the ball $B(r)$ of radius
$r$ using the Gauss theorem.

Since $R\gg\sigma$, the particles are statistically likely to be
outside of the effective range of the interaction potential $\phi(r)$
for most of the time. Therefore, we can reasonably regard them as
being independent, that is,
\begin{equation}
\delta\BV U(\BV x,\BV y)=\BV U(\BV y,\BV x)-\BV U(\BV x,\BV y)=\BV
u(\BV y)-\BV u(\BV x),\qquad\nabla_{\BV y}\cdot\delta\BV U(\BV x,\BV
y)=\nabla\cdot\BV u(\BV y),
\end{equation}
where $\BV u(\BV x)$ is the usual single-particle average velocity
from \eqref{eq:rho_u_p}. Further, in the synthetic scenario of
Section~\ref{sec:mean_distance}, it was presumed that $\nabla\cdot\BV
u$ is constant inside $B(R)$. Therefore, we can factor $\nabla\cdot\BV
u$ out of the integral, which yields
\begin{equation}
\frac 1V\int_0^R\left(\int_{B(r)}\nabla_{\BV z}\cdot\delta\BV U(\BV x,
\BV x+\BV z)\dif\BV z \right)\dif r=\frac{\nabla\cdot\BV u}V\int_0^R
\left(\int_{B(r)}\dif\BV z\right)\dif r=\frac 14\left(\frac{3V}{4\pi}
\right)^{1/3}\nabla\cdot\BV u.
\end{equation}
As we can see, the integral above is the exact match for
\eqref{eq:DF_divu}.  Therefore, we introduce the following correction
of the conventional pair correlation function in
\eqref{eq:conventional_collision}:
\begin{equation}
\label{eq:novel_collision}
g=g_0\,\exp\left(\frac 1{2\theta}(\BV w-\BV v)\cdot\delta\BV U(\BV x,
\BV y)\right).
\end{equation}

\subsection{The corrected collision integral and the novel transport
  equations}
\label{sec:novel_equations}

Above, we used the assumption of a constant temperature solely for
convenience in deriving the average velocity correction for the pair
correlation function in \eqref{eq:novel_collision}. Henceforth, we
assume that the density $\rho$, velocity $\BV u$ and temperature
$\theta$ are all variables.  Substituting~\eqref{eq:novel_collision}
into the collision integral \eqref{eq:f_collision}, we obtain
\begin{equation}
\label{eq:collision2}
\coll[f]=-\frac 1mf(\BV x,\BV v)\int_{\RR^3}\int_{B(\sigma)}\parderiv{
  \phi(\|\BV z\|)}{\BV z}e^{-\frac{\phi(\|\BV y\|)}\theta+\frac 1{2
    \theta}(\BV w-\BV v)\cdot\delta\BV U(\BV x,\BV x+\BV z)}Y(\|\BV z
\|) f(\BV x+\BV z,\BV w)\dif\BV z\dif\BV w.
\end{equation}
For a short-range $\phi(r)$, \eqref{eq:collision2} becomes, in the
constant-density hydrodynamic limit \cite{Abr17},
\begin{subequations}
\begin{equation}
\label{eq:collision}
\coll[f]=\frac 1\rho\parderiv{\bar\phi}{\BV x}f(\BV x,\BV v)+\BM
J\int_{\RR^3}(\BV w-\BV v)f(\BV x,\BV v)f(\BV x,\BV w)\dif\BV w,
\end{equation}
\begin{equation}
\label{eq:bphi}
\bar\phi=\frac{2\pi\rho p}{3m}\int_0^\sigma\Big(1-e^{-\frac{\phi(r)}
  \theta}\Big)\parderiv{}r\left(r^3Y(r)\right)\dif r,
\end{equation}
\begin{equation}
\label{eq:J}
\BM J=-\frac 1{2m\theta}\int_0^\sigma e^{-\frac{\phi(r)}\theta} Y(r)
\phi'(r)\left(\int_{\mathbb S_1}\BV n\delta\BV U(\BV x,\BV x+r\BV
n)^T\dif\BV n\right)r^2\dif r,
\end{equation}
\end{subequations}
where $\bar\phi$ is the mean field potential
\cite{Abr22,Abr23,Abr24,Abr25,Abr26,Abr27} (a.k.a.~the second
coefficient of the virial expansion, or the van der Waals effect). The
computation of the collision integral in~\eqref{eq:collision} is shown
in Appendices~\ref{sec:standard_collision}
and~\ref{sec:novel_collision}. The collision moment of
\eqref{eq:collision} for the momentum equation is
\begin{equation}
\langle\BM I\cdot\rangle_{\coll[f]}=\frac 1\rho\parderiv{\bar\phi}{\BV
  x}\int_{\RR^3}f(\BV x,\BV v)\dif\BV v+\BM J\int_{\RR^6}(\BV w-\BV v)
f(\BV x,\BV v)f(\BV x,\BV w)\dif\BV v\dif\BV w=\nabla\bar\phi.
\end{equation}
The integral of the $\BM J$-dependent part above is zero, because it
is skew-symmetric with respect to renaming $\BV v\leftrightarrow\BV
w$.  The collision moment of \eqref{eq:collision} for the pressure
equation is computed as follows:
\begin{multline}
\langle(\BV v-\BV u)\cdot\rangle_{\coll[f]}=\frac 1\rho\parderiv{\bar
  \phi}{\BV x}\cdot\int_{\RR^3}(\BV v-\BV u)f(\BV x,\BV v)\dif\BV v+
\int_{\RR^6}(\BV v-\BV u)^T\BM J(\BV w-\BV v)\\f(\BV x,\BV v)f(\BV
x,\BV w)\dif\BV v\dif\BV w=\BM J:\int_{\RR^6}(\BV v-\BV u)(\BV w-\BV
v)^Tf(\BV x,\BV v)f(\BV x,\BV w)\dif\BV v\dif\BV w.
\end{multline}
The integral over the velocities alone is
\begin{multline}
\int_{\RR^6}(\BV v-\BV u)(\BV w-\BV v)^Tf(\BV x,\BV v)f(\BV x,\BV w)
\dif\BV v\dif\BV w=\int_{\RR^6}(\BV v-\BV u)(\BV w-\BV u)^Tf(\BV x,
\BV v)f(\BV x,\BV w)\dif\BV v\dif\BV w\\-\int_{\RR^6}(\BV v-\BV u)^2
f(\BV x,\BV v)f(\BV x,\BV w)\dif\BV v\dif\BV w=-\rho p\BM I,
\end{multline}
which results in
\begin{equation}
\langle(\BV v-\BV u)\cdot\rangle_{\coll[f]}=-\rho p\trace(\BM J).
\end{equation}
Substituting the collision moments, computed above, into the transport
equations for the density, momentum and pressure
in~\eqref{eq:rho_u_p_coll} yields
\begin{subequations}
\label{eq:rho_u_p_transport}
\begin{equation}
\label{eq:rho_u_transport}
\Dderiv\rho t+\rho\nabla\cdot\BV u=0,\qquad\rho\Dderiv{\BV u}t+\nabla
p+\nabla\cdot\BV\Sigma=-\nabla\bar\phi,
\end{equation}
\begin{equation}
\label{eq:p_transport}
\Dderiv pt+ \frac 53p\nabla\cdot\BV u+\frac 23(\BV\Sigma:\nabla\BV u+
\nabla\cdot\BV q)=\frac 23\rho p\trace(\BM J).
\end{equation}
\end{subequations}
As we can see, the density and momentum equation are unchanged from
our previous works \cite{Abr22,Abr23,Abr24,Abr25,Abr26,Abr27}, and the
effect of the novel correction \eqref{eq:novel_collision} to the pair
correlation function manifests solely in the pressure equation.

\section{A closure for \texorpdfstring{$\trace(\BM J)$}{tr(J)} and the
damped pressure equation}
\label{sec:trJ_closure}

In order to be able to solve the system of equations in
\eqref{eq:rho_u_p_transport}, we need to construct an estimate of
$\trace(\BM J)$, given by \eqref{eq:J}, via $\rho$, $\BV u$ and $p$,
defined in \eqref{eq:rho_u_p}. Unfortunately, for the purpose of
computation of $\trace(\BM J)$, we cannot assume that the two
particles are statistically independent (as we did above in
Section~\ref{sec:correction} to validate \eqref{eq:novel_collision} in
the synthetic scenario of Section~\ref{sec:mean_distance}), because,
according to the integral in \eqref{eq:J}, the two particles are
within the range of the potential and thus interact. Therefore, we
have to resort to a different, yet similarly crude, approach. As we
will see below, it nonetheless leads to realistic results, and is
therefore worth looking into.

Computing the trace of \eqref{eq:J}, we have
\begin{multline}
\trace(\BM J)=-\frac 1{2m\theta}\int_0^\sigma e^{-\frac{\phi(r)}
  \theta}Y(r)\phi'(r)\left(\int_{\mathbb S_1}\BV n\cdot\delta\BV U(\BV
x,\BV x+r\BV n) r^2\dif\BV n\right)\dif r\\=-\frac 1{2m\theta}\int_0^
\sigma e^{-\frac{\phi(r)}\theta}Y(r)\phi'(r)\left(\int_{\mathbb S (r)}
\BV n\cdot\delta\BV U(\BV x,\BV x+r\BV n)\dif A\right)\dif r\\=-\frac
1{2m\theta}\int_0^\sigma e^{-\frac{\phi(r)}\theta}Y(r)\phi'(r)\left(
\int_{B(r)}\nabla_{\BV z}\cdot\delta\BV U(\BV x,\BV x+\BV z)\dif\BV z
\right)\dif r,
\end{multline}
where we used the Gauss theorem in the last line to switch from the
surface integral over the sphere $\mathbb S(r)$ to the volume integral
over the ball $B(r)$. For the convenience of remaining calculations,
we will further assume that $\phi(r)$, while remaining smooth,
approaches the hard sphere potential of radius $\sigma$, that is,
\begin{equation}
\label{eq:phi_HS}
\phi(r)\to\left\{\begin{array}{l@{\qquad}l} \infty, & r\leq\sigma,
\\ 0, & r>\sigma.\end{array}\right.
\end{equation}
This simplifies the double integral for $\trace(\BM J)$ above into a
single volume integral:
\begin{equation}
\label{eq:trJ_HS_2}
\trace(\BM J)=\frac{Y(\sigma)}{2m}\int_{B(\sigma)}\nabla_{\BV z}\cdot
\delta\BV U(\BV x,\BV x+\BV z)\dif\BV z.
\end{equation}
The next step is to evaluate the integral above, for which we need to
estimate
\begin{equation}
\nabla_{\BV y}\cdot\delta\BV U(\BV x,\BV y)=\nabla_{\BV y}\cdot\big(
\BV U(\BV y,\BV x)-\BV U(\BV x,\BV y)\big).
\end{equation}
The first term above, that is, $\nabla_{\BV y}\cdot\BV U(\BV y,\BV
x)$, has ``affinity'' to the single-particle divergence
$\nabla\cdot\BV u(\BV y)$, as shown in Section~\ref{sec:correction}.
On the other hand, the second term $\nabla_{\BV y}\cdot\BV U(\BV x,\BV
y)$ is more mysterious, because the differentiation is conducted with
respect to the coordinate of the second particle, and it is unclear
(at least intuitively) what this means.

To estimate the term $\nabla_{\BV y}\cdot\delta\BV U(\BV x,\BV y)$, we
introduce the function $\BV U_*(\BV p,\BV q)$, defined as
\begin{equation}
\label{eq:u_*}
\BV U_*\left(\frac{\BV x+\BV y}2,\BV x-\BV y
\right)= \BV U(\BV x,\BV y),\qquad\|\BV x-\BV y\|\leq\sigma.
\end{equation}
In other words, within the range of the potential interaction, $\BV
U_*$ happens to be $\BV U$ itself, but expressed as a function of the
midpoint between $\BV x$ and $\BV y$, and their difference.
In terms of $\BV U_*$, $\nabla_{\BV y}\cdot\BV U(\BV x,\BV y)$ and
$\nabla_{\BV y}\cdot\BV U(\BV y,\BV x)$ are given via
\begin{subequations}
\begin{equation}
\label{eq:uyx}
\nabla_{\BV y}\cdot\BV U(\BV y,\BV x)=\frac 12\nabla_{\BV p}\cdot\BV
U_*\left(\frac{\BV x+\BV y}2,\BV y-\BV x\right)+\nabla_{\BV q}\cdot\BV
U_*\left(\frac{\BV x+\BV y}2,\BV y-\BV x\right),
\end{equation}
\begin{equation}
\nabla_{\BV y}\cdot\BV U(\BV x,\BV y)=\frac 12\nabla_{\BV p}\cdot\BV
U_*\left(\frac{\BV x+\BV y}2,\BV x-\BV y\right)-\nabla_{\BV q}\cdot\BV
U_*\left(\frac{\BV x+\BV y}2,\BV x-\BV y\right),
\end{equation}
\end{subequations}
where $\nabla_{\BV p}$ and $\nabla_{\BV q}$ denote the differentiation
in the first and second argument of $\BV U_*$, respectively.
Therefore, the integrand of \eqref{eq:trJ_HS_2} is given via
\begin{multline}
\nabla_{\BV z}\cdot\delta\BV U(\BV x,\BV x+\BV z)=\frac 12\nabla_{
  \BV p}\cdot\left[\BV U_*\left(\BV x+\frac{\BV z}2,\BV z\right)-\BV
  U_*\left(\BV x+\frac{\BV z}2,-\BV z\right)\right]\\+\nabla_{\BV q}
\cdot\left[\BV U_*\left(\BV x+ \frac{\BV z}2,\BV z\right)+\BV U_*
  \left(\BV x+\frac{\BV z}2,-\BV z\right)\right],
\end{multline}
Now we need to compute the integral in \eqref{eq:trJ_HS_2}, which is
\begin{multline}
\int_{B(\sigma)}\nabla_{\BV z}\cdot\delta\BV U(\BV x,\BV x+\BV z)\dif
\BV z=\frac 12\int_{B(\sigma)}\nabla_{\BV p}\cdot\left[\BV U_*\left(
  \BV x-\frac{\BV z}2,-\BV z\right)-\BV U_*\left(\BV x+\frac{\BV z}2,
  -\BV z\right)\right]\dif\BV z\\+\int_{B(\sigma)}\nabla_{\BV q}\cdot
\left[\BV U_*\left(\BV x-\frac{\BV z}2,-\BV z\right)+\BV U_*\left(\BV
  x+\frac{\BV z}2,-\BV z\right)\right]\dif\BV z\\=2\int_{B(\sigma)
}\nabla_{\BV q}\cdot\BV U_*\left(\BV x-\frac{\BV z}2,-\BV z\right)\dif
\BV z+O(\sigma^4),
\end{multline}
where we changed the dummy variable of integration $\BV z\to-\BV z$ in
the first terms under both integrals, and isolated the
$O(\sigma)$-variations in the first argument of $\BV U_*$ into a
separate higher-order term. The expression for $\trace(\BM J)$ in
\eqref{eq:trJ_HS_2} is, therefore, given via
\begin{equation}
\label{eq:trJ_HS_leading_order}
\trace(\BM J)=\frac{Y(\sigma)}m\int_{B(\sigma)}\nabla_{\BV q}\cdot\BV
U_*\left(\BV x-\frac{\BV z}2,-\BV z\right)\dif\BV z+O(\sigma^4/m),
\end{equation}
where the trailing higher-order term vanishes in the hydrodynamic
limit. It turns out that, in the leading order, $\trace(\BM J)$
consists only of the second term in \eqref{eq:uyx}. Therefore, we need
to think of a way to remove the first term from the right-hand side of
\eqref{eq:uyx}.

Recall that, in the two-particle collision integrals of
Appendix~\ref{sec:collision}, the dependence of involved quantities on
the difference between particle locations (the second argument of $\BV
U_*$) is much more sensitive to changes, than their dependence on the
midpoint (the first argument of $\BV U_*$). This happens due to the
involvement of the potential $\phi(r)$, which depends on the
difference between particle locations.  Here, we assume that $\BV
U_*(\BV p,\BV q)$ depends on its two arguments in a similar manner ---
it is much more sensitive to the changes in $\BV q$ than in $\BV p$.
If so, then $\nabla_{\BV q}\cdot\BV U_*$ oscillates much more rapidly
than $\nabla_{\BV p}\cdot\BV U_*$, as the locations of particles
change over time. Subsequently, we introduce an ansatz that a running
time average $\langle\cdot\rangle$, with a suitable averaging window,
filters out $\nabla_{\BV q}\cdot\BV U_*$, while not affecting
$\nabla_{\BV p}\cdot\BV U_*$. The application of such a running time
average to \eqref{eq:uyx} deletes the second term from the right-hand
side, while leaving the first term intact:
\begin{equation}
\langle\nabla_{\BV y}\cdot\BV U(\BV y,\BV x)\rangle=\frac 12\nabla_{
  \BV p}\cdot\BV U_*\left(\frac{\BV x+\BV y}2,\BV y-\BV x\right).
\end{equation}
We now express
\begin{subequations}
\begin{equation}
\nabla_{\BV q}\cdot\BV U_*\left(\frac{\BV x+\BV y}2,\BV y-\BV x\right)
=\nabla_{\BV y}\cdot\BV U(\BV y,\BV x)-\langle\nabla_{\BV y}\cdot\BV
U(\BV y,\BV x)\rangle,\quad\text{and thus}
\end{equation}
\begin{equation}
\nabla_{\BV q}\cdot\BV U_*\left(\BV x-\frac{\BV z}2,-\BV z\right)
=\nabla_1\cdot\BV U(\BV x,\BV x-\BV z)-\langle\nabla_1\cdot\BV U(\BV
x,\BV x-\BV z)\rangle,
\end{equation}
\end{subequations}
where ``$\nabla_1$'' denotes the differentiation with respect to the
first argument of $\BV U(\BV x,\BV y)$, irrespectively of the
variables present in both arguments.  The expression for $\trace(\BM
J)$ in~\eqref{eq:trJ_HS_leading_order} becomes
\begin{equation}
\label{eq:trJ_2}
\trace(\BM J)=\frac{Y(\sigma)}m\int_{B(\sigma)}[\nabla_1\cdot\BV U(\BV
  x,\BV x-\BV z)-\langle\nabla_1\cdot\BV U(\BV x,\BV x-\BV z)\rangle]
\dif\BV z.
\end{equation}
By definition, spatial averages over the second particle yield the
corresponding single-particle quantities, which means that
\begin{equation}
\int_{B(\sigma)}\nabla_1\cdot\BV U(\BV x,\BV x-\BV z)\dif\BV z=\frac{4
  \pi}3\sigma^3\nabla\cdot\BV u,\quad\int_{B(\sigma)}\langle\nabla_1
\cdot\BV U(\BV x,\BV x-\BV z)\rangle\dif\BV z=\frac{4\pi}3\sigma^3
\langle\nabla\cdot\BV u\rangle.
\end{equation}
Substituting the expressions above into \eqref{eq:trJ_2} yields
\begin{equation}
\label{eq:trJ_closure}
\trace(\BM J)=\frac{8Y(\sigma)}{\rho_\HS}(\nabla\cdot\BV u-\langle
\nabla\cdot\BV u\rangle),\qquad\text{where}\qquad\rho_\HS=\frac{6m}{
  \pi\sigma^3}
\end{equation}
is the density of the hard-sphere particle of the mass $m$ and
diameter $\sigma$.
Substituting the above expression for $\trace(\BM J)$ into the
pressure equation \eqref{eq:p_transport}, we arrive at
\begin{equation}
\Dderiv pt+\frac 53\left(1-\frac{16}5\frac{\rho Y(\sigma)}{\rho_\HS}
\right)p\nabla\cdot\BV u+\frac 23(\BV\Sigma:\nabla\BV u+\nabla\cdot
\BV q)=-\frac{16}3\frac{\rho Y(\sigma)}{\rho_\HS}p\langle\nabla\cdot
\BV u\rangle.
\end{equation}
We recall that, first, $Y(\sigma)=1+O(\rho/\rho_\HS)$ \cite{Bou86},
and, second, $\rho/\rho_\HS\sim 6.5\cdot 10^{-4}\ll 1$ at normal
conditions (see \cite{Abr26}, also refer to Table~\ref{tab:ref_param}
below for details).  Therefore, we discard all
$O(\rho/\rho_\HS)$-corrections to unity in the pressure equation
above.  This transforms the system of equations for $\rho$, $\BV u$
and $p$ in \eqref{eq:rho_u_p_transport} into
\begin{subequations}
\label{eq:rho_u_p_HS_divU}
\begin{equation}
\label{eq:rho_u_HS_divU}
\Dderiv\rho t+\rho\nabla\cdot\BV u=0,\qquad\rho\Dderiv{\BV u}t+\nabla
\left[p\left(1+\frac{4\rho}{\rho_\HS}\right)\right]+\nabla\cdot\BV
\Sigma=\BV 0,
\end{equation}
\begin{equation}
\label{eq:p_HS_divU}
\Dderiv pt+\frac 53p\nabla\cdot\BV u+\frac 23(\BV\Sigma:\nabla\BV u+
\nabla\cdot\BV q)=-\frac{16}3\frac{\rho p}{\rho_\HS}\langle\nabla
\cdot\BV u\rangle,
\end{equation}
\end{subequations}
where the mean field potential $\bar\phi$ in the momentum equation has
been replaced with its hard-sphere formula \cite{Abr24}, which is
easily computable from \eqref{eq:bphi} by substituting
\eqref{eq:phi_HS}.

It remains to determine a tractable closure for $\langle\nabla\cdot\BV
u\rangle$ in the right-hand side of~\eqref{eq:p_HS_divU} via the
variables $\rho$ and $p$. For that, we note that, once such closure
has been achieved, the variables of the novel system in
\eqref{eq:rho_u_p_HS_divU} model the ``slow'' components of gas
dynamics which we observe at low Mach numbers, while the ``fast''
adiabatic fluctuations around them are described by the standard
compressible Navier--Stokes equations (that is,
\eqref{eq:rho_u_p_HS_divU} without the van der Waals effect in the
momentum equation, and the $\langle\nabla\cdot\BV u\rangle$-term in
the pressure equation). Therefore, we can write the momentum equation
via
\begin{equation}
\label{eq:momentum_slow_fast}
\Dderiv{\BV u}t=-\frac{\nabla p}\rho+\text{rapid fluctuations},
\end{equation}
where $\BV u$ is the ``total'' velocity of gas which involves both
slow and fast components of dynamics, while $\rho$ and $p$ are the
``slow'' density and pressure variables of \eqref{eq:rho_u_p_HS_divU},
and the term $\nabla p/\rho$ can be treated as a slow forcing, applied
to rapid adiabatic fluctuations.

Next, let us consider the dynamics where the slow term $\nabla p/\rho$
above is absent, and only the fast adiabatic motions are present. In
such a scenario, the generic solution for $\BV u$ is an acoustic wave,
given via
\begin{equation}
\BV u\sim e^{-at}\cos(2\pi\nu t),
\end{equation}
where $\nu$ is the frequency of the acoustic wave, and $a$ is the
attenuation coefficient due to the presence of dissipative effects.
Subsequently, the kinetic energy of the wave behaves as
\begin{equation}
\|\BV u\|^2\sim e^{-2at}\cos^2(2\pi\nu t)=\frac 12e^{-2at}(1+\cos(4
\pi\nu t)).
\end{equation}
Further, we assume that the oscillations occur on a much shorter time
scale than the attenuation, such that the running time average filters
out the former, but not the latter. This means that, along the stream
line, $\langle\|\BV u\|^2\rangle$ should decay as
\begin{equation}
\langle\|\BV u\|^2\rangle\sim e^{-2at}.
\end{equation}
For the decay of the running average of $\BV u$ itself, we estimate
\begin{equation}
\langle\BV u\rangle\sim\sqrt{\langle\|\BV u\|^2\rangle}\sim e^{-at}.
\end{equation}
The above estimate can be used to approximate $\langle\BV u\rangle$ in
\eqref{eq:momentum_slow_fast} as a linear response to the slow forcing
$\nabla p/\rho$ by means of the Green--Kubo formula
\cite{Gre,Kub1,Kub2}:
\begin{equation}
\label{eq:Green_Kubo}
\langle\BV u\rangle=-\left(\int_0^\infty e^{-at}\dif t\right)
\frac{\nabla p}\rho=-\frac 1a\frac{\nabla p}\rho.
\end{equation}
In Appendix~\ref{sec:attenuation}, we compute the attenuation
coefficient $a$ in the same fashion as done in the Stokes--Kirchhoff
law of sound attenuation \cite{Stokes1845,Kirchhoff1868}. It turns out
that, while $a$ is indeed a scalar coefficient in the Fourier space,
in the physical space it becomes a differential operator:
\begin{equation}
\label{eq:attenuation}
a=\frac 1{3\rho}\left(\frac{32\alpha\sigma_{SB}T_0^3}{5R}-\beta\mu
\Delta\right),\qquad\beta=2+\frac 1\Pran,\qquad\Pran=\frac 52\frac{R
  \mu}\kappa.
\end{equation}
Above, $\Pran$ is the monatomic Prandtl number. The physical
parameters, used above, are provided in Table~\ref{tab:ref_param}.
\begin{table}
\begin{tabular}{|c|c|c|}
\hline
Parameters & Symbol & Value \\
\hline\hline
Stefan--Boltzmann constant & $\sigma_{SB}$ & $5.67\cdot 10^{-8}$ kg
s$^{-3}$ K$^{-4}$ \\
Gas constant, air & $R$ & $287$ m$^2$ s$^{-2}$ K$^{-1}$ \\
Viscosity, air & $\mu$ & $1.825\cdot 10^{-5}$ kg m$^{-1}$ s$^{-1}$ \\
Prandtl number, air & $\Pran$ & $0.7$ \\
Absorption coefficient, air & $\alpha$ & $5.5\cdot 10^{-5}$ m$^{-1}$ \\
Pressure, sea level & $p_0$ & $1.013\cdot 10^5$ kg m$^{-1}$ s$^{-2}$ \\
Temperature, sea level & $T_0$ & $293.15$ K \\
Density, sea level & $\rho_0$ & $1.204$ kg m$^{-3}$ \\
Hard sphere density & $\rho_\HS$ & $1850$ kg m$^{-3}$ \cite{Abr23} \\
\hline
\end{tabular}
\vspace{0.5EM}
\caption{Reference values of physical parameters used throughout the
  work.}
\label{tab:ref_param}
\end{table}
Finally, substituting \eqref{eq:attenuation} into
\eqref{eq:Green_Kubo}, and applying the divergence operator to both
sides, we arrive at the equation for $\langle\nabla\cdot\BV u\rangle$:
\begin{equation}
\frac 13\left(-\frac{32\alpha\sigma_{SB}T_0^3}{5R}+\beta\mu\Delta
\right)\langle\nabla\cdot\BV u\rangle=\Delta p.
\end{equation}
The transport equations in \eqref{eq:rho_u_p_HS_divU} become
\begin{subequations}
\label{eq:rho_u_p_HS}
\begin{equation}
\label{eq:rho_u_HS}
\Dderiv\rho t+\rho\nabla\cdot\BV u=0,\qquad\rho\Dderiv{\BV u}t
+\nabla\left[p\left(1+\frac{4\rho}{\rho_\HS}\right)\right]+
\nabla\cdot\BV\Sigma =\BV 0,
\end{equation}
\begin{equation}
\label{eq:p_HS}
\Dderiv pt+\frac 53p\nabla\cdot\BV u+\frac 23(\BV\Sigma:\nabla\BV u+
\nabla\cdot\BV q)=\frac{16\rho p}{\rho_\HS}\left(\frac{32\alpha
    \sigma_{SB}T_0^3}{5R}-\beta\mu\Delta
  \right)^{-1}\Delta p.
\end{equation}
\end{subequations}
In our past works \cite{Abr22,Abr23,Abr24,Abr26,Abr27}, we studied the
model of gas flow with the density and momentum transport equations
given by \eqref{eq:rho_u_HS}, with the pressure being either preserved
along the stream lines (balanced flow), or set to a constant
throughout the domain (inertial flow). Such behavior is observed in
very large scale flows (e.g.~the geostrophic flow in the Earth
atmosphere \cite{Cha}). In contrast, here the pressure equation
\eqref{eq:p_HS} has a dissipative term in the right-hand side, which
manifests as diffusion at large scales (due to radiative cooling), and
as linear damping at small scales (due to viscous effects). This
appears to correspond to the observed behavior of pressure --- at
large scales, regions of different pressure appear to ``redistribute''
themselves into more even patterns, whereas at small scales pressure
nonuniformities simply ``vanish'' into the local background state.

The switching between linear damping and diffusion occurs at the
spatial scale
\begin{equation}
\label{eq:L_transition}
L=\left(\frac{5R\beta\mu}{32\alpha \sigma_{SB} T_0^3}\right)^{1/2}.
\end{equation}
For the standard values of parameters above (refer to
Table~\ref{tab:ref_param}), $L\approx 6$ meters.

\section{Some properties of the damped pressure equation}
\label{sec:properties}

\subsection{Linear wave structure in a resting gas}

Here, we examine the linear wave solutions of \eqref{eq:rho_u_p_HS}
when the gas is at rest, or is moving at a uniform velocity, such that
an appropriate Galilean shift can be made. For that, we follow the
same procedure as in Appendix~\ref{sec:attenuation}, which results in
the nondimensional linear equations of the form
\begin{subequations}
\begin{equation}
\parderiv{\tilde\rho'}{\tilde t}+\tilde\chi=0,\qquad\parderiv{\tilde
  \chi}{\tilde t}+\frac{p_0t_0^2}{\rho_0L^2}\tilde\Delta\left(\tilde
p'+4\eta\tilde\rho'\right)=\frac 43\frac{\mu t_0}{\rho_0L^2}\tilde
\Delta\tilde\chi,
\end{equation}
\begin{equation}
\parderiv{\tilde p'}{\tilde t}+\frac 53\tilde\chi=\frac{5t_0}{3\rho_0}
\left(\frac\mu{\Pran L^2} \tilde\Delta-\frac {32}5\frac{\alpha
  \sigma_{SB}T_0^3}R\right)(\tilde p'-\tilde\rho')+\frac{16\eta p_0t_0
}{L^2}\left(\frac{32\alpha \sigma_{SB}T_0^3}{5R}-\frac{\beta\mu}{L^2}
  \tilde\Delta \right)^{-1}\tilde\Delta\tilde p',
\end{equation}
\begin{equation}
\text{where}\qquad\eta=\frac{\rho_0}{\rho_\HS},\quad\BV{\tilde x}=\frac{\BV x}L,\quad\tilde t=\frac t{
  t_0},\quad\tilde\rho'=\frac\rho{\rho_0}-1,\quad\tilde\chi=t_0
\nabla\cdot\BV u,\quad\tilde p'=\frac p{p_0}-1.
\end{equation}
\end{subequations}
From Table~\ref{tab:ref_param}, it follows that the packing fraction
$\eta\approx 6.5\cdot 10^{-4}$ (also see \cite{Abr26}).

Here, we choose the reference spatial scale $L$ as in
\eqref{eq:L_transition}.  Additionally, we define
\begin{equation}
t_0=\frac{\beta\mu}{16\eta p_0}\approx 6\cdot 10^{-7}\text{
  seconds},\qquad\varepsilon=\left(\frac{2\alpha \sigma_{SB}
  T_0^4\mu}{3\eta p_0^2} \right)^{1/2}\approx 2.1\cdot 10^{-7},
\end{equation}
which results in
\begin{subequations}
\begin{equation}
\parderiv{\tilde\rho'}{\tilde t}+\tilde\chi=0,\qquad\parderiv{\tilde
  \chi}{\tilde t}+\frac{3\beta\varepsilon^2}{80\eta}\tilde\Delta(
\tilde p'+4\eta\tilde\rho')=\frac 45\varepsilon^2\tilde\Delta\tilde
\chi,
\end{equation}
\begin{equation}
\parderiv{\tilde p'}{\tilde t}+\frac 53\tilde\chi=\varepsilon^2\left(
\frac 1\Pran\tilde\Delta-\beta\right)(\tilde p'-\tilde\rho')+(1-
\tilde\Delta)^{-1}\tilde\Delta\tilde p'.
\end{equation}
\end{subequations}
In the Fourier space, the above system of PDE becomes a linear system
of ODE:
\begin{subequations}
\label{eq:nondim_waves}
\begin{equation}
\deriv{\hat\rho'}{\tilde t}=-\hat\chi,\qquad\deriv{\hat\chi}{\tilde t}
=\frac{3\beta\varepsilon^2}{80\eta}\|\BV k\|^2\left(\hat p'+4\eta\hat
\rho'\right)-\frac 45\varepsilon^2\|\BV k\|^2\hat\chi,
\end{equation}
\begin{equation}
\deriv{\hat p'}{\tilde t}=-\frac 53\hat\chi+\varepsilon^2\left(\beta+
\frac{\|\BV k\|^2}\Pran\right)\hat\rho'-\left[\frac{\|\BV k\|^2}{1+\|
    \BV k\|^2}+\varepsilon^2\left(\beta+\frac{\|\BV k\|^2}\Pran\right)
  \right]\hat p'.
\end{equation}
\end{subequations}
In Appendix~\ref{sec:waves} we find that the eigenvalues of
\eqref{eq:nondim_waves} are given via
\begin{equation}
\label{eq:lambda_waves}
\lambda_0=-\frac{\|\BV k\|^2}{1+\|\BV k\|^2}+O(\varepsilon^2), \qquad
\lambda_{1,2}=-\frac{\varepsilon^2\beta}{32\eta}(1+\|\BV k\|^2)\pm
\frac{i\varepsilon}2\sqrt{\frac{3\beta}5}\|\BV k\|+O(\varepsilon^3),
\end{equation}
while the eigenvectors $\BV e_0$ and $\BV e_{1,2}$ project onto the
pressure and density fluctuations, respectively. As we can see, on the
time scale of $L$ (that is, meters) the pressure fluctuations exhibit
a very rapid decay towards the background pressure state on the time
scale of $O(t_0)$, whereas the density fluctuations exhibit
oscillations on the $O(t_0/\varepsilon)$ time scale, combined with the
decay on the $O(t_0/\varepsilon^2)$ time scale.

It is not difficult to verify by direct substitution that, in the
leading order, $\hat\rho'$ satisfies the second-order equation
\begin{equation}
\deriv{^2\hat\rho'}{\tilde t^2}+\frac{\varepsilon^2\beta}{16\eta}(1+\|
\BV k\|^2)\deriv{\hat\rho'}{\tilde t}+\frac{3\varepsilon^2\beta}{20}
\|\BV k\|^2\hat\rho'=0,
\end{equation}
which in the physical space translates into
\begin{subequations}
\label{eq:density_waves}
\begin{equation}
\parderiv{^2\rho'}{t^2}+\left(\frac 1{\tau_\rho}-D_\rho\Delta\right)
\parderiv{\rho'}t=c_\rho^2\Delta\rho',\qquad\tau_\rho=\frac{16\eta t_0
}{\varepsilon^2\beta}=\frac{3\rho_0p_0}{2 \alpha\sigma_{SB}T_0^4
  \rho_\HS}\approx 1.2\text{ hours},
\end{equation}
\begin{equation}
D_\rho=\frac{\varepsilon^2 \beta L^2}{16\eta t_0}=\frac{5\beta\mu
  \rho_\HS}{48\rho_0^2}\approx 8.3\cdot 10^{-3}\;\frac{\mathrm m^2}{
  \mathrm s},\qquad c_\rho=\sqrt{\frac{3\beta}5}\frac{\varepsilon
  L}{2t_0}=2\sqrt{\frac{p_0}{\rho_\HS}}\approx 15\;\frac{\mathrm m}{
  \mathrm s}.
\end{equation}
\end{subequations}
As we can see, the density fluctuations $\rho'$ around its background
state $\rho_0$ solve a damped wave equation with the phase speed of
the waves given by $c_\rho\sim 15$ m$/$s at sea level. We also observe
that, since the pressure is equilibrated, the temperature fluctuations
should exhibit qualitatively same motions (i.e.~the density waves can
also be detected as the ``thermal'' waves)

It is worth noting that, at the upper end of the troposphere, $p_0$
decreases roughly by an order of magnitude, from which it follows that
$c_\rho$ should become $\sim 5$ m$/$s.  Anecdotally, planetary
atmospheric waves in the equatorial zone, e.g.~the equatorial Kelvin
and Rossby waves, as well as the Madden--Julian oscillation (MJO),
tend to have observed phase speeds in the range 5--15 m/s. However, as
far as we know, the van der Waals effect is not included in the models
of such waves; for example, none of the four perspective models of the
MJO, reviewed in \cite{ZhaAdaKhoWanYan}, uses the van der Waals effect
to explain the phase speed. Given that the van der Waals effect is
present in reality, and the phase speed of the density waves it
produces roughly corresponds to the observed speed of the MJO and
other equatorial waves, this issue appears to merit further
investigation.

\subsection{Diagnostic pressure approximation}

As was found above, on the spatial scale of meters, the pressure
variable $p$ in \eqref{eq:p_HS} is damped by a combination of linear
damping and diffusion on a very short time scale ($\sim 10^{-6}$
seconds). We can use this observation to simplify the pressure
equation using the averaging approximation of multiscale dynamics
\cite{PavStu}. For that, we nondimensionalize the variables in
\eqref{eq:rho_u_p_HS} similarly to how it was done in the previous
section, except that the equations remain nonlinear. Namely, we
introduce
\begin{equation}
\BV{\tilde x}=\frac{\BV x}L,\qquad\tilde t=\frac t{t_0},\qquad \tilde
\rho=\frac\rho{\rho_0},\qquad\BV{\tilde u}=\frac{t_0}L\BV u,\qquad
\tilde p=\frac p{p_0}.
\end{equation}
Here, however, the reference time $t_0$ is set to $L/U$, where $U$ is
the reference speed variation of the flow, while both $L$ and $U$
remain parameters. In addition, we introduce the Mach number $\Mach$,
the Reynolds number $\Rey$, and the monatomic Knudsen number $\Knud$,
respectively, via
\begin{equation}
\Mach=U\sqrt{\frac{3\rho_0}{5p_0}},\qquad\Rey=\frac{\rho_0UL}\mu,
\qquad\Knud=\frac\mfp L=\sqrt{\frac{5\pi}6}\frac\Mach\Rey,
\end{equation}
where $\mfp$ is the mean free path between molecular collisions. In
the nondimensional variables, with the Newton and Fourier laws
\eqref{eq:Newton_Fourier} in place, the equations
\eqref{eq:rho_u_p_HS} become
\begin{subequations}
\begin{equation}
\Dderiv{\tilde\rho}{\tilde t}+\tilde\rho\tilde\nabla\cdot\BV{\tilde u}
=0,\qquad\tilde\rho\Dderiv{\BV{\tilde u}}{\tilde t}+\frac 3{5\Mach^2}
\nabla[\tilde p(1+4\eta\tilde\rho)]+\frac 1\Rey\tilde\nabla\cdot\BV{
  \tilde\Sigma}=\BV 0,
\end{equation}
\begin{equation}
\Dderiv{\tilde p}{\tilde t}+\frac 53\tilde p\tilde\nabla\cdot\BV{
  \tilde u}+\frac{10}{9\Rey}\left[\Mach^2\BV{\tilde\Sigma}:\tilde
  \nabla\BV{\tilde u}+\frac 3{2\Pran}\tilde\Delta\left(\frac{\tilde p
  }{\tilde\rho}\right)\right]=\frac{3\eta\Rey}{\Mach^2}\tilde\rho
\tilde p\left(\frac{\pi\alpha\sigma_{SB}T_0^4\mu}{p_0^2\Knud^2}-
\frac{5\beta}{16}\tilde\Delta \right)^{-1}\tilde\Delta\tilde p,
\end{equation}
\begin{equation}
\BV{\tilde\Sigma}=-\tilde\nabla\BV{\tilde u}-\tilde\nabla\BV{\tilde u
}^T+\frac 23(\tilde\nabla\cdot\BV{\tilde u})\BM I.
\end{equation}
\end{subequations}
Next, introduce the pressure deviation $\tilde p'$ via
\begin{equation}
\tilde p=1+\frac{5\Mach^2}3\tilde p',
\end{equation}
where the constant scaling coefficient in front of $\tilde p'$ is
chosen for convenience. Let us now assume that, due to the strong
pressure damping, the nondimensional pressure deviation $|\tilde
p'|\ll 1$. Then, we can linearize with respect to the nondimensional
pressure, and discard the $O(\eta)$-terms in comparison to unity:
\begin{subequations}
\begin{equation}
\label{eq:rho_u_nondim}
\Dderiv{\tilde\rho}{\tilde t}+\tilde\rho\tilde\nabla\cdot\BV{\tilde u}
=0,\qquad\tilde\rho\Dderiv{\BV{\tilde u}}{\tilde t}+\nabla\tilde p'+
\frac{12\eta}{5\Mach^2}\tilde\nabla\tilde\rho+\frac 1\Rey\tilde\nabla
\cdot\BV{\tilde\Sigma}=\BV 0,
\end{equation}
\begin{equation}
\Mach^2\Dderiv{\tilde p'}{\tilde t}+\tilde\nabla\cdot\BV{\tilde u}+
\frac 1\Rey\left[\frac 23\Mach^2\BV{\tilde\Sigma}:\tilde\nabla\BV{
    \tilde u}+\frac 1\Pran\tilde\Delta\left(\frac{\tilde p}{\tilde
    \rho}\right)\right]=3\eta\Rey\tilde\rho\left(\frac{\pi\alpha
  \sigma_{SB}T_0^4\mu}{p_0^2\Knud^2}-\frac{5\beta}{16}\tilde\Delta
\right)^{-1}\tilde\Delta\tilde p'.
\end{equation}
\end{subequations}
Here, we choose the reference values for the Mach and Reynolds numbers
as $\Mach\lesssim 0.1$ (a typical subsonic flow), $\Rey\gtrsim 10^3$
(turbulent regime). As we can see, the advective derivative in the
pressure equation is scaled by the small parameter $\Mach^2$, which
means that $\tilde p'$ is the fast variable. Therefore, according to
the averaging formalism \cite{PavStu}, we can replace $\tilde p'$ in
the momentum equation with its average from the pressure equation
(with $\tilde\rho$ and $\BV{\tilde u}$ being fixed parameters), which
gives an approximation for the density and momentum dynamics on the
time scale of $O(1)$, in the nondimensional units. However, due to the
linearity in $\tilde p'$, the invariant measure for the pressure
equation is singular, and fixed at the steady state of $\tilde p'$ for
the given $\tilde\rho$ and $\BV{\tilde u}$. Additionally, we discard
the $O(\Rey^{-1})$-terms from this steady state, which yields
\begin{subequations}
\begin{equation}
\label{eq:p_nondim}
\tilde\Delta\tilde p'=\frac 1{3\eta\Rey}\left(\frac{\pi\alpha\sigma_{
    SB}T_0^4\mu}{p_0^2\Knud^2}-\frac{5\beta}{16}\tilde\Delta \right)
\left(\frac{\tilde\nabla\cdot\BV{\tilde u}}{\tilde\rho}\right),\qquad
\text{or}
\end{equation}
\begin{equation}
\label{eq:tau_zeta}
\Delta p=\frac\rho\tau\nabla\cdot\BV u-\Delta\left(\zeta\nabla\cdot
\BV u\right),\qquad\tau=\frac{3R\rho^2}{2\alpha\sigma_{SB}T_0^3
  \rho_\HS},\qquad\zeta=\frac{5\beta\mu\rho_\HS}{48\rho}.
\end{equation}
\end{subequations}
Here, observe that the diagnostic pressure equation above in
\eqref{eq:tau_zeta} includes both the linear damping term (with the
decay time $\tau$), and the diffusion term, with the diffusion
coefficient $\zeta$. Each of these two terms can be dominant,
depending on the spatial scale. In particular, if the spatial scale is
much greater than the transition scale in \eqref{eq:L_transition}
(i.e.~$\gg 6$ meters at normal conditions), then the linear damping
term dominates, and \eqref{eq:tau_zeta} effectively becomes the
``weakly compressible'' pressure equation from our recent work
\cite{Abr25}. Conversely, at the spatial scales much shorter than
\eqref{eq:L_transition}, the diffusion term dominates, and we can
formally ``undo'' the Laplacians on both sides of \eqref{eq:tau_zeta},
thereby expressing the pressure gradient for the momentum equation
\eqref{eq:rho_u_HS} directly via the velocity divergence:
\begin{equation}
\rho\Dderiv{\BV u}t+\frac{4p}{\rho_\HS}\nabla\rho=\mu\left(\Delta\BV
u+\frac 13\nabla(\nabla\cdot\BV u)\right)+\left(1+\frac{4\rho}{
  \rho_\HS}\right)\nabla(\zeta\nabla\cdot\BV u).
\end{equation}
Observing that $\rho/\rho_\HS\ll 1$, and $|p-p_0|/p_0\sim \Mach^2/\eta
\Rey\ll 1$, we set $p\to p_0$ in the van der Waals term in left-hand
side, and discard the $O(\rho/\rho_\HS)$-correction to unity in the
right-hand side. This yields the closed momentum equation in the form
\begin{equation}
\label{eq:rho_u_WC}
\rho\Dderiv{\BV u}t+\frac{4p_0}{\rho_\HS}\nabla\rho=\mu\Delta\BV u+
\nabla\left[\left(\zeta+\frac\mu 3\right)\nabla\cdot\BV u\right].
\end{equation}
The momentum equation above in \eqref{eq:rho_u_WC} is very similar to
the one for the inertial flow we used in
\cite{Abr22,Abr23,Abr24,Abr26,Abr27}, with the exception of the
additional diffusive term $\nabla((\zeta+\mu/3)\nabla\cdot\BV u)$.

Remarkably, the diffusion coefficient $\zeta$, defined in
\eqref{eq:tau_zeta}, matches the description of the {\em second
  viscosity}, which acts selectively on the divergence of velocity
$\nabla\cdot\BV u$, and does not affect vorticity $\nabla\times\BV u$
(see, for example, Section 81 of \cite{LanLif}). At normal conditions,
we estimated $\rho/\rho_\HS\approx 6.5\cdot 10^{-4}$ (see
Table~\ref{tab:ref_param}, also \cite{Abr23,Abr26}).  This means that
$\zeta/\mu\sim 550$, which agrees with some
observations~\cite{Cramer2012} to an order of magnitude.

Lastly, we verify that the system consisting of the density and
momentum equations in \eqref{eq:rho_u_HS}, and the pressure equation
in \eqref{eq:tau_zeta}, captures the density waves from
\eqref{eq:density_waves}. For that, we write the density and momentum
transport equations as
\begin{subequations}
\begin{equation}
\parderiv\rho t+\nabla\cdot(\rho\BV u)=0,
\end{equation}
\begin{equation}
\parderiv{(\rho\BV u)}t+\nabla\cdot(\rho\BV u^2)+\nabla p+\frac{4p_0}{
  \rho_\HS}\nabla\rho=\mu\left(\Delta\BV u+\frac 13\nabla(\nabla\cdot
\BV u)\right).
\end{equation}
\end{subequations}
Computing the time derivative of the density equation, the divergence
of the momentum equation, equating the mixed derivatives, discarding
the $O(\BV u^2)$-term, replacing the advective derivative with the
partial time derivative in the density equation, and substituting
$\Delta p$ from \eqref{eq:tau_zeta} yields the closed equation for
$\rho$ alone:
\begin{equation}
\parderiv{^2\rho}{t^2}+\frac 1\tau\parderiv\rho t-\Delta\left[\left(
  \zeta+\frac 43\mu\right)\frac 1\rho\parderiv\rho t\right]=\frac{4p_0
}{\rho_\HS}\Delta\rho.
\end{equation}
A linearization of the above equation around $\rho_0$ matches
\eqref{eq:density_waves} with the exception of the $\mu$-term, which
is an $O(\eta)$-correction to $\zeta$.

\subsection{Linearization around a two-dimensional shear flow with
constant vorticity}

In our work~\cite{Abr27}, we studied the behavior of linearized
transport equations for the two-dimensional inertial flow, comprised
of the density equation in \eqref{eq:rho_u_HS}, and the momentum
equation being essentially same as in \eqref{eq:rho_u_WC}, but without
the second viscosity $\zeta$. The linearization of the solution was
computed around the steady state with a constant density, and a linear
shear velocity profile (or constant vorticity). Here, we conduct the
same analysis as in \cite{Abr27}, but with the system of equations,
comprised of the density equation in \eqref{eq:rho_u_HS}, and the
novel momentum equation in \eqref{eq:rho_u_WC}, which now includes the
second viscosity $\zeta$.

Following \cite{Abr27}, we assume that the flow is two-dimensional,
and use the Helmholtz decomposition of the velocity field into the
stream function $\psi$ and the potential $\varphi$,
\begin{equation}
\BV u=\nabla^\perp\psi+\nabla\varphi,\quad\nabla^\perp
=\begin{pmatrix} -\partial /\partial y \\ \partial/\partial
x\end{pmatrix},\quad\chi=\nabla\cdot\BV u=\Delta\varphi,\quad
\omega=\nabla^\perp\cdot \BV u=\Delta\psi,
\end{equation}
where $\nabla^\perp$ is the two-dimensional curl operator, and we
introduced separate notations for the velocity divergence $\chi$ and
the two-dimensional vorticity $\omega$, respectively. Applying the
divergence and curl to the nondimensional momentum equation in
\eqref{eq:rho_u_nondim}, we obtain, in the nondimensional variables,
\begin{subequations}
\begin{equation}
\Dderiv{\tilde\rho}{\tilde t}=-\tilde\rho\tilde\chi,\qquad\Dderiv{
  \tilde\omega}{\tilde t}+\tilde\omega\tilde\chi=\tilde\nabla^\perp
\cdot\bigg[\frac{\tilde\nabla}{\tilde\rho}\left(\frac{4\tilde\chi
  }{3\Rey}-\tilde p'\right)+\frac{\tilde \nabla^\perp\tilde\omega
  }{\Rey\,\tilde\rho}\bigg],
\end{equation}
\begin{equation}
\Dderiv{\tilde\chi}{\tilde t}+\|\tilde\nabla(\tilde\nabla^\perp\tilde
\psi+\tilde\nabla\tilde\varphi)\|_F^2-\tilde\omega^2=\tilde\nabla\cdot
\bigg[\frac{\tilde\nabla}{\tilde\rho}\left(\frac{4\tilde\chi}{3\Rey}-
  \frac{12\eta\tilde\rho}{5\Mach^2}-\tilde p'\right)+\frac{\tilde
    \nabla^\perp\tilde\omega}{\Rey\,\tilde\rho}\bigg],
\end{equation}
\end{subequations}
where $\|\cdot\|_F$ denotes the Frobenius norm. Next, we linearize the
equations above around the steady state with constant density and
vorticity (which corresponds to a linear shear velocity), given via
\begin{equation}
\tilde\rho_0=\tilde\omega_0=1,\qquad\tilde\psi_0=\tilde y/2,\qquad
\tilde\chi_0=\tilde \varphi_0=0.
\end{equation}
For small fluctuations $\tilde\rho'$, $\tilde\psi'$ and
$\tilde\omega'$ around this steady state, the linearized density,
vorticity and divergence equations are
\begin{subequations}
\begin{equation}
\parderiv{\tilde\rho'}{\tilde t}-\tilde y\parderiv{\tilde\rho'}{
  \tilde x}=-\tilde\chi,\qquad\parderiv{\tilde\omega'}{\tilde t}-
\tilde y\parderiv{\tilde\omega'}{\tilde x}=\frac 1\Rey\tilde\Delta
\tilde\omega'-\tilde\chi,
\end{equation}
\begin{equation}
\label{eq:chi_nondim}
\parderiv{\tilde\chi}{\tilde t}-\tilde y\parderiv{\tilde\chi}{\tilde x
}=2\left(\parderiv{^2\tilde\varphi}{\tilde x\partial\tilde y}+
\parderiv{^2\tilde\psi'}{\tilde x^2}\right)-\frac{\vartheta\Rey}{
  \Mach^2}\tilde\chi+\left(\frac 1{\Rey_\zeta}+\frac 4{3\Rey}\right)
\tilde \Delta\tilde\chi-\frac{\beta\Rey_\zeta}{4\Rey\Mach^2} \tilde
\Delta\tilde\rho',
\end{equation}
\end{subequations}
where we substituted $\tilde\Delta\tilde p'$ from \eqref{eq:p_nondim},
and defined the ``second Reynolds number'' $\Rey_\zeta$, and a
nondimensional parameter $\vartheta$, respectively, via
\begin{equation}
\label{eq:Re2}
\Rey_\zeta=\frac{48\eta}{5\beta}\Rey,\qquad\vartheta=\frac{2\alpha
  \sigma_{SB} T_0^4\mu\rho_\HS}{5\rho_0p_0^2}\sim 2.5\cdot 10^{-14}.
\end{equation}
Observing that $\Rey_\zeta/\Rey\sim\eta\sim 10^{-3}$, we can discard
the term $4/3\Rey$ in the coefficient of $\tilde\Delta\tilde\chi$
above. Following \cite{Abr27}, we transform the system above into the
Fourier space:
\begin{subequations}
\begin{equation}
\parderiv{\hat\rho'}{\tilde t}+k_x\parderiv{\hat\rho'}{k_y}=-\hat\chi,
\qquad\parderiv{\hat\omega'}{\tilde t}+k_x\parderiv{\hat\omega'}{k_y}
=-\frac{\|\BV k\|^2}\Rey\hat\omega'-\hat\chi,
\end{equation}
\begin{equation}
\parderiv{\hat\chi}{\tilde t}+k_x\parderiv{\hat\chi}{k_y}=\bigg(\frac{
  2k_xk_y}{\|\BV k\|^2}-\frac{\vartheta\Rey}{\Mach^2}-\frac{\|\BV k\|^
  2}{\Rey_\zeta}\bigg)\hat\chi+\frac{2k_x^2}{\|\BV k\|^2}\hat\omega'+
\frac{\beta\Rey_\zeta}{4\Rey}\frac{\|\BV k\|^2}{\Mach^2}\hat\rho',
\end{equation}
\begin{equation}
\text{where we expressed}\qquad\hat\varphi=-\frac{\hat\chi}{\|\BV k\|
  ^2},\qquad \hat\psi'=-\frac{\hat\omega'}{\|\BV k\|^2}.
\end{equation}
\end{subequations}
As in \cite{Abr27}, we convert the system of partial differential
equations (PDE) above into a system of linear ordinary differential
equations (ODE) on the characteristic straight lines in the $(\tilde
t,k_y)$-plane, with $k_x$ being a fixed parameter, which specifies the
slope of the corresponding characteristic.  The characteristics are
given by
\begin{equation}
\big(\tilde t,k_y(\tilde t)\big)=(0,k_{y,0})+\tilde t(1,k_x).
\end{equation}
On these characteristics, the system of PDE above becomes a system of
linear ODE:
\begin{subequations}
\label{eq:Fourier_space}
\begin{equation}
\deriv{}{\tilde t}\begin{pmatrix}\hat\rho' \\ \hat\omega' \\ \hat\chi
\end{pmatrix}=\BM A(\tilde t)\begin{pmatrix}\hat\rho' \\ \hat\omega' \\
\hat\chi\end{pmatrix},\qquad\BM A(\tilde t)=\begin{pmatrix} 0 & 0 & -1 \\
0 & -\frac{\|\BV k(\tilde t)\|^2}\Rey & -1 \\ \frac{\beta\Rey_\zeta}{4\Rey}
\frac{\|\BV
k(\tilde t)\|^2}{\Mach^2} & \frac{2k_x^2}{\|\BV k(\tilde t)\|^2} &
\frac{2k_xk_y(\tilde t)}{\|\BV k(\tilde t)\|^2}-\frac{\vartheta\Rey
}{\Mach^2}-\frac{\|\BV k(\tilde t)\|^2}{\Rey_\zeta}
\end{pmatrix},
\end{equation}
\begin{equation}
\text{where}\qquad k_y(\tilde t)=k_{y,0}+k_x\tilde t,\qquad\|\BV
k(\tilde t)\|^2=k_x^2+k_y^2(\tilde t).
\end{equation}
\end{subequations}
In the current setting, the two Reynolds numbers $\Rey_\zeta$ and
$\Rey$ define, respectively, the large and small scales of the flow.
Therefore, below we examine the eigenvalues of the matrix $\BM A$ in
\eqref{eq:Fourier_space} for two different scenarios: $\|\BV
k\|^2\lesssim\Rey_\zeta$ (large scale dynamics), and
$\Rey_\zeta\ll\|\BV k\|^2\lesssim\Rey$ (small scale dynamics).
Additionally, we examine the asymptotic behavior of
\eqref{eq:Fourier_space} as $t\to\infty$.

\subsection{Large scale dynamics and the critical value of the Reynolds
  number}

Here, we examine the eigenvalues of the matrix $\BM A$ in
\eqref{eq:Fourier_space} for $\|\BV k\|^2\lesssim\Rey_\zeta$, that is,
at the large scales. In Appendix~\ref{sec:large_scale}, we find
\begin{subequations}
\begin{equation}
\label{eq:l0_ls}
\lambda_0=-\frac{\|\BV k\|^2}\Rey\frac 1{1+\frac{5\Mach^2k_x^2}{6\eta
    \|\BV k\|^4}}+O(\eta^2),
\end{equation}
\begin{equation}
\label{eq:l12_ls}
\lambda_{1,2}=\frac{k_xk_y}{\|\BV k\|^2}-\frac{\vartheta\Rey}{2\Mach^2
}-\frac{\|\BV k\|^2}{2\Rey_\zeta}\pm i\sqrt{\frac{12\eta\|\BV k\|^2}{5
    \Mach^2}+\frac{2k_x^2}{\|\BV k\|^2}-\left(\frac{k_xk_y}{\|\BV k
    \|^2}-\frac{\vartheta\Rey}{2\Mach^2}-\frac{\|\BV k\|^2}{2
    \Rey_\zeta}\right)^2}+O(\eta).
\end{equation}
\end{subequations}
It turns out that the eigenvalue $\lambda_0$ in \eqref{eq:l0_ls} is
always real and negative. Generally, $\lambda_{1,2}$ in
\eqref{eq:l12_ls} can be both real and complex-conjugate, however, in
Appendix~\ref{sec:large_scale} we show that real $\lambda_{1,2}$ are
always negative (while we are naturally interested more in positive
eigenvalues, due to the associated instability). A positive real part
in~\eqref{eq:l12_ls} implies that
\begin{equation}
\label{eq:instability_region}
\frac{2k_xk_y}{\|\BV k\|^2}>\frac{\|\BV k\|^2}{\Rey_\zeta}+ \frac{
  \vartheta\Rey}{\Mach^2},
\end{equation}
for which the expression under the radical is guaranteed to be
positive. 

Remarkably, the origin of the term $k_xk_y/\|\BV k\|^2$, which causes
the instability in the real part of \eqref{eq:l12_ls}, can be traced
back to $\partial^2\tilde\varphi/\partial\tilde x\partial\tilde y$ in
\eqref{eq:chi_nondim}, which, in turn, is the result of coupling of
the small scale divergence with the large scale vorticity. At the same
time, the imaginary part of \eqref{eq:l12_ls} in the leading order
represents a linear wave with the phase speed of
$\sqrt{12\eta/5\Mach^2}$, which in the dimensional units translates
into $\sqrt{4p_0/\rho_\HS}$ (i.e. it is the same density wave as in
\eqref{eq:density_waves}, which is caused by the van der Waals
effect). Just as we observed in our work \cite{Abr27}, the turbulent
dynamics are the result of the exponentially growing fluctuations of
the flow in the presence of the instability
\eqref{eq:instability_region}, further mixed by the density waves due
to the van der Waals effect. The main difference between \cite{Abr27}
and the current work is the presence of the second viscosity $\zeta$
in the latter.

\begin{figure}%
\includegraphics[width=0.5\textwidth]{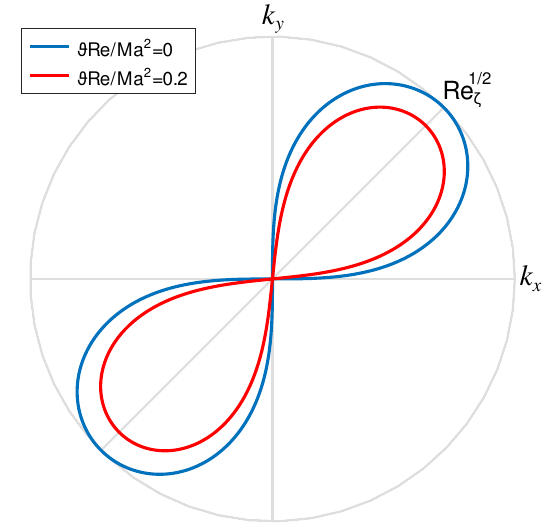}%
\caption{The instability region corresponding to
  \eqref{eq:instability_region}.  For illustration, two sample regions
  are shown: one corresponds to $\vartheta\Rey/\Mach^2=0$, and another
  to $\vartheta\Rey/\Mach^2=0.2$.}%
\label{fig:instability_region}%
\end{figure}

In the polar coordinates $(\|\BV k\|,\alpha)$ of the
$(k_x,k_y)$-plane, \eqref{eq:instability_region} is expressed in a
somewhat more convenient form:
\begin{equation}
\|\BV k\|\leq\sqrt{\Rey_\zeta\left(\sin 2\alpha-\frac{\vartheta\Rey}{
    \Mach^2}\right)}.
\end{equation}
The instability condition \eqref{eq:instability_region} holds inside
the two petal-shaped regions in the first and third quadrants of the
$(k_x,k_y)$-plane, depicted in Figure~\ref{fig:instability_region}.
The points of the maximal distance from the origin lie on the
45-degree line at the distance of
$\sqrt{\Rey_\zeta(1-\vartheta\Rey/\Mach^2)}$.  Here, recall from
\eqref{eq:Re2} that $\vartheta\sim 10^{-14}$, and, therefore, we need
$\Rey/\Mach^2\sim 10^{14}$ for the radiative cooling effect to
suppress the instability. At $\Mach\sim 0.1$, this requires $\Rey\sim
10^{12}$, that is, the stabilizing effect of the radiative cooling
practically manifests at spatial scales of thousands of kilometers,
and we can ignore it at smaller scales.

Observe that the characteristics of \eqref{eq:Fourier_space} are
parallel vertical lines in the $(k_x,k_y)$-plane. The solutions
propagate upward in the right-hand half of the plane, and downward in
the left-hand half of the plane. Therefore, in the linearized system
\eqref{eq:Fourier_space}, instability develops at large scales ($\|\BV
k\|^2\lesssim\Rey_\zeta$) when the solution passes through the
instability region in Figure~\ref{fig:instability_region}, and then
decays as the solution continues to move toward the small scales
($\Rey_\zeta\ll\|\BV k\|^2\lesssim\Rey$). This is known as the {\em
  direct cascade}. In our work \cite{Abr27}, we suggested that, in
order for an {\em inverse cascade} to be created (that is, for
turbulent fluctuations to propagate from small scales to large
scales), one likely needs a flow with two adjacent regions with
vorticity of opposite signs, such as the Poiseuille flow or a jet.

Remarkably, the instability condition \eqref{eq:instability_region}
imposes restrictions on the representative size of the flow relative
to the second Reynolds number $\Rey_\zeta$. Indeed, if $\Rey_\zeta$ is
``too small'', then the instability region in
Figure~\ref{fig:instability_region} may not be able to accommodate the
smallest wavenumbers which are present in the flow, and, as a result,
the flow would remain stable. As an example, consider a typical
axisymmetric flow, such as a jet. In its longitudinal section, there
are two shear flow regions, which are the mirror symmetries of each
other about the axis of the flow. Since the width of each shear flow
region is half that of the jet, the wavenumbers with $\|\BV k\|<2$
correspond to the spatial scales greater than the representative size
of each region. Therefore, in order for the flow to be unstable, the
instability region in \eqref{eq:instability_region} must accommodate
the wavenumbers with $\|\BV k\|>2$, which corresponds to the critical
value of the Reynolds number
\begin{equation}
\Rey_\zeta\sim\|\BV k\|^2\sim 4,\qquad\text{or}\qquad\Rey=\frac{5\beta
}{48\eta}\Rey_\zeta\sim\frac{5\beta}{12\eta}.
\end{equation}
From Table~\ref{tab:ref_param}, it follows that the critical value of
the Reynolds number is $\Rey\sim 2200$. This estimate matches the
well-known Reynolds criterion \cite{Rey83,Rey,Let,Men} to an order of
magnitude.

\subsection{Small scale dynamics and the fluctuation decay rate}

Here, we examine the eigenvalues of the matrix $\BM A$ in
\eqref{eq:Fourier_space} at small scales, that is, $\Rey_\zeta\ll\|\BV
k\|^2\lesssim\Rey$. In Appendix~\ref{sec:small_scale}, we find that
all three roots are real and negative:
\begin{equation}
\label{eq:l_ss}
\lambda_0=-\frac{12\eta\Rey_\zeta}{5\Mach^2}+O(\eta),\qquad\lambda_1
=-\frac{\|\BV k\|^2}\Rey+O(\eta),\qquad\lambda_2=-\frac{\|\BV k\|^2
}{\Rey_\zeta}+O(1).
\end{equation}
The roots $\lambda_1$ and $\lambda_2$ obviously correspond to the
diagonal entries of $\BM A$, and scale with $\|\BV k\|^2$ (viscous
decay). The root $\lambda_0$, on the other hand, is constant and
corresponds to a linear damping. Further,
\begin{equation}
\lambda_0>\lambda_1\qquad\text{once}\qquad\|\BV k\|>\frac{24\eta\Rey}{
  5\sqrt\beta\Mach},
\end{equation}
and the dominant dissipative behavior of solutions of
\eqref{eq:Fourier_space} switches from the viscous diffusion to a
linear damping. In the dimensional variables, $\lambda_0$ has
the units of inverse time, which means it can be computed via
\begin{equation}
\lambda_0=-\frac{12\eta\Rey_\zeta}{5\Mach^2}\frac UL=-\frac{192}{5
  \beta}\frac{\rho^2}{\rho_\HS^2}\frac p\mu.
\end{equation}

\subsection{Asymptotic stability}

We have to note that the eigenvalues of $\BM A$, calculated above, at
best provide crude estimates of the behavior of solutions of
\eqref{eq:Fourier_space}. The reason is that the system in
\eqref{eq:Fourier_space} is non-autonomous, and the ``instantaneous''
properties of $\BM A(\tilde t)$ at a given time $\tilde t$ do not
generally extend onto the qualitative behavior of solutions over
periods of time (there is a counter-example due to Markus and Yamabe,
see \cite{MarYam}, p.~310).

However, the asymptotic behavior can still be quantified for those
non-autonomous systems, for which $\BM A(\tilde t)$ loses dependence
on $\tilde t$ as $\tilde t\to\infty$ (or a suitable change of the time
variable exists, for which the preceding holds). In our case, observe
that the largest entries of $\BM A(\tilde t)$ in
\eqref{eq:Fourier_space} behave as $O(\tilde t^2)$ as $\tilde
t\to\infty$.  Therefore, a cubic change of the time variable leads to
a vanishing time-depending part of \eqref{eq:Fourier_space} at
infinite times, so that the system becomes ``asymptotically
autonomous''. Further, the matrix of the resulting autonomous system
has distinct eigenvalues. As a result, Levinson's theorem
\cite{CodLev} applies, and in Appendix~\ref{sec:asymptotic} we use it
to obtain the following asymptotic solution of
\eqref{eq:Fourier_space}:
\begin{equation}
\label{eq:l_asymptotic}
\begin{pmatrix}\tilde\rho \\ \tilde\omega \\ \tilde\chi\end{pmatrix}
\sim C_0e^{-\frac{12\eta\Rey_\zeta\tilde t}{5\Mach^2}}\begin{pmatrix}1
  \\ 0 \\ \frac{12\eta\Rey_\zeta}{5\Mach^2} \end{pmatrix}+ C_1
e^{-\frac{k_x^2\tilde t^3}{3\Rey}}\begin{pmatrix}0 \\ 1 \\ 0 \end{pmatrix}
+C_2e^{-\frac{k_x^2\tilde t^3}{3\Rey_\zeta}}\begin{pmatrix} 0 \\ 0 \\ 1
\end{pmatrix}.
\end{equation}
It turns out that \eqref{eq:Fourier_space} is asymptotically stable.
The linear damping as the leading order behavior is due to the lack of
dissipation in the density equation (it is easy to check that a
viscous term in the density equation leads to the viscous diffusion in
all eigenvectors).

\section{Conclusions}

The behavior of pressure in a low Mach, high Reynolds number gas flow
is a long-standing mystery. The traditional molecular-kinetic model,
which consists of the Boltzmann equation and the resulting Euler or
Navier--Stokes equations, predicts adiabatic flow where the gas
compresses with increasing temperature. However, in reality we observe
that, at low Mach numbers, the pressure becomes stabilized, which
instead results in the expansion of the gas when its temperature
increases. At the same time, at a high Mach number, the gas indeed
behaves as predicted by the Euler or Navier--Stokes equations, that
is, it compresses when heated, forming acoustic waves, shock
transitions, and other relevant high-speed phenomena. It is clear that
the macroscopic thermodynamic behavior of a real gas is affected by
the Mach number of the flow.

In the current work, we propose a molecular-kinetic hypothesis which
seems to explain such a mysterious behavior of pressure. Our reasoning
is simple: if the gas expands, then the average distance between
particles must increase, and vice versa if it compresses. Yet, the
conventional BBGKY closure, based on the equilibrium Gibbs state of
the system, does not account for this effect; instead, the average
distance between two generic particles remains unaffected by the
expansion or compression of the gas. To ameliorate this discrepancy,
we modify the pair correlation function of the BBGKY closure of the
deterministic, Vlasov-type collision integral, so that it correctly
models the rate of change of the average distance between particles
depending on the macroscopic compression or expansion rate of the
gas. Remarkably, we find that our correction of the pair correlation
function leaves the density and momentum transport equations
unchanged, and manifests only as an additional term in the pressure
equation.

For the novel pressure equation, we propose a closure based on the
Green--Kubo linear response formula, which relates the additional term
to the pressure gradient via the attenuation coefficient of an
acoustic wave. It results in a pressure dissipation effect, which
combines viscous diffusion at large scales, and linear damping at
small scales. At normal conditions, the acoustic waves become
suppressed by this dissipation at relevant spatial scales, with the
density (or ``thermal'') waves emerging in their stead due to the van
der Waals effect. Anecdotally, the speed of propagation of these
density/thermal waves roughly matches that of atmospheric equatorial
planetary waves, such as the Madden--Julian oscillation.

The dissipative effect in the pressure equation becomes stronger at
low Mach numbers, which results in the pressure variable being
approximated by its own steady state, for a given density and velocity
variables. This, in turn, leads to the momentum equation which has a
novel dissipative effect acting selectively on the divergence of
velocity, while the vorticity of the flow remains unaffected. This
dissipative effect combines linear damping at large scales (similarly
to the empirical model we studied in \cite{Abr25}) with diffusion at
small scales. If the latter dominates the former (which happens at
length scales of less than a meter), the pressure equation effectively
becomes algebraic, and, as a result, the pressure gradient in the
momentum equation is replaced with the gradient of the velocity
divergence, scaled by the second viscosity. Our hypothesis calculates
the second viscosity by dividing the usual shear viscosity by a
coefficient proportional to the packing fraction. This results in the
value of the second viscosity exceeding that of the shear viscosity by
a factor of roughly five hundred, at normal conditions.

Finally, we analyze the new momentum equation in the same setting as
we did in \cite{Abr27} for the inertial flow. It turns out that the
linear instability, which generates turbulent dynamics, is now
governed by the second viscosity. Since the latter is roughly three
orders of magnitude greater than the shear viscosity, the
corresponding critical value of the Reynolds number exceeds unity by
the same factor, which agrees with observations and experiments.
Contrary to the general understanding, the critical value of the
Reynolds number appears to be unrelated directly to the shear
viscosity in the momentum equation; it, however, emerges as a
consequence of the dissipative pressure dynamics, which, at low Mach
numbers, manifest in the momentum equation in the form of the second
viscosity.

We note that in a laminar flow (which occurs below the critical value
of the Reynolds number), the velocity divergence is suppressed roughly
five hundred times stronger than the vorticity, which results in the
overall dynamics being qualitatively similar to those of the
incompressible flow (where the velocity divergence is set to
zero). This could be the reason why the incompressible Euler and
Navier--Stokes equations happen to be satisfactory models of
compressible dilute gases in laminar flows at low Mach numbers.

\ack The work was supported by the Simons Foundation grant \#636144.

\appendix

\allowdisplaybreaks

\section{Derivation of the density, momentum and pressure equations}
\label{sec:moment_derivation}

\subsection{The density equation}

For the zero-order velocity moment, we obtain, from
\eqref{eq:m_transport},
\begin{equation}
\parderiv{\langle 1\rangle_f}t+\nabla\cdot\langle\BV v\rangle_f=-\left
\langle\parderiv 1{\BV v}\cdot\right\rangle_{\coll[f]}.
\end{equation}
From \eqref{eq:rho_u_p} we recall that $\langle 1\rangle_f=\rho$,
$\langle\BV v\rangle_f=\rho\BV u$, and, observing that the right-hand
side is zero, we arrive at the density transport equation in the form
\begin{equation}
\parderiv\rho t+\nabla\cdot(\rho\BV u)=0,\qquad\text{or}\qquad
\Dderiv\rho t+\rho\nabla\cdot\BV u=0.
\end{equation}

\subsection{The momentum equation}

For the first-order moment, we obtain, from \eqref{eq:m_transport},
\begin{equation}
\parderiv{\langle\BV v\rangle_f}t+\nabla\cdot\langle\BV v^2\rangle_f
=-\left\langle\parderiv{\BV v}{\BV v}\cdot\right\rangle_{\coll[f]}.
\end{equation}
Here, we first express
\begin{equation}
\label{eq:2nd_moment}
\langle\BV v^2\rangle_f=\langle(\BV v-\BV u)^2\rangle_f+\BV u\langle
\BV v\rangle_f+\langle\BV v\rangle_f\BV u-\langle 1\rangle_f\BV u^2=
\BM P+\rho\BV u^2,
\end{equation}
where $\BM P=\langle(\BV v-\BV u)^2\rangle_f$ is the pressure tensor.
Observing that $\partial\BV v/\partial\BV v=\BM I$, we arrive at
\begin{equation}
\parderiv{(\rho\BV u)}t+\nabla\cdot(\rho\BV u^2)+\nabla\cdot\BM P
=-\langle\BM I\cdot\rangle_{\coll[f]},
\end{equation}
or, after subtracting the density equation, multiplied by $\BV u$,
\begin{equation}
\rho\Dderiv{\BV u}t+\nabla\cdot\BM P=-\langle\BM I\cdot
\rangle_{\coll[f]}.
\end{equation}
Here, we split the pressure tensor $\BM P$ into the sum of its own
trace (which is the pressure $p$), and the remainder, which is called
the stress, and is denoted by $\BV\Sigma$ \eqref{eq:stress_heat_flux}:
\begin{equation}
\label{eq:pressure_stress}
\BM P=p\BM I+\BV\Sigma,\qquad p=\frac 13 \trace(\BM P).
\end{equation}
This yields the momentum equation in the form
\begin{equation}
\rho\Dderiv{\BV u}t+\nabla p+\nabla\cdot\BV\Sigma=-\langle\BM I\cdot
\rangle_{\coll[f]}.
\end{equation}

\subsection{The pressure equation}

For the second-order moment, we obtain, from \eqref{eq:m_transport},
\begin{equation}
\parderiv{\langle\|\BV v\|^2\rangle_f}t+\nabla\cdot\langle\|\BV v\|^2
\BV v\rangle_f=-\left\langle\parderiv{\|\BV v\|^2}{\BV v}\cdot\right
\rangle_{\coll[f]}.
\end{equation}
Here, we express, from \eqref{eq:2nd_moment} and
\eqref{eq:pressure_stress},
\begin{equation}
\langle\|\BV v\|^2\rangle_f=\langle\|\BV v-\BV u\|^2\rangle_f+\rho
\|\BV u\|^2=3p+\rho\|\BV u\|^2,
\end{equation}
\begin{multline}
\langle\|\BV v\|^2\BV v\rangle_f=\langle\|\BV v\|^2(\BV v-\BV u)
\rangle_f+\langle\|\BV v\|^2\rangle_f\BV u=\langle(\|\BV v-\BV u\|^2+2
\BV u\cdot\BV v-\|\BV u\|^2)(\BV v-\BV u)\rangle_f\\+(3p+\rho\|\BV
u\|^2)\BV u=\langle\|\BV v-\BV u\|^2(\BV v-\BV u)\rangle_f+2\langle
(\BV v-\BV u)\BV v\rangle_f\BV u+(3p+\rho\|\BV u\|^2)\BV u\\= 2(\BV
q+\BM P\BV u)+(3p+\rho\|\BV u\|^2)\BV u=2(\BV q+\BV\Sigma\BV u)+(5p+
\rho\|\BV u\|^2)\BV u,
\end{multline}
where the heat flux $\BV q$ is defined in \eqref{eq:stress_heat_flux}.
Observing that $\partial\|\BV v\|^2/\partial\BV v=2\BV v$, we arrive
at
\begin{equation}
\parderiv{}t\left(3p+\rho\|\BV u\|^2\right)+\nabla\cdot\left((5p+\rho
\|\BV u\|^2)\BV u+2(\BV q+\BV\Sigma\BV u)\right)=-2\left\langle\BV
v\cdot\right \rangle_{\coll[f]}.
\end{equation}
Next, we express from the density and momentum equations,
\begin{multline}
\parderiv{(\rho\|\BV u\|^2)}t=2\BV u\cdot\parderiv{(\rho\BV u)}t-
\parderiv\rho t\|\BV u\|^2\\=\|\BV u\|^2\nabla\cdot(\rho\BV u)-2\BV
u\cdot\left(\nabla\cdot(\rho\BV u^2)+ \nabla p+\nabla\cdot\BV\Sigma
+\langle\BM I\cdot\rangle_{\coll[f]}\right),
\end{multline}
which yields, upon substitution,
\begin{equation}
\parderiv pt+\nabla\cdot(p\BV u)+\frac 23(p\nabla\cdot\BV
u+\BV\Sigma:\nabla\BV u+\nabla\cdot\BV q)=-\frac 23
\left\langle(\BV v-\BV u)\cdot\right\rangle_{\coll[f]},
\end{equation}
or, with the use of the advective derivative,
\begin{equation}
\Dderiv pt+\frac 53p\nabla\cdot\BV u+\frac 23(\BV\Sigma:\nabla\BV u+
\nabla\cdot\BV q)=-\frac 23\left\langle(\BV v-\BV u)\cdot\right
\rangle_{\coll[f]}.
\end{equation}

\section{Computation of the collision integral and its moments}
\label{sec:collision}

Here we present the computation of the collision integral in
\eqref{eq:collision2} in the hydrodynamic limit for a short-range
potential $\phi(r)$, with the range $\sigma$. Noting that the
integrand of \eqref{eq:collision2} is nonzero only within the
effective range of the potential, we can expand the pair correlation
function in powers of $\BV y$ as follows:
\begin{multline}
\coll[f]=-\frac 1m\int_{\RR^3}\int_{B(\sigma)}\parderiv{\phi(\|\BV z\|
  )}{\BV z}e^{-\frac{\phi(\|\BV z\|)}\theta}Y(\|\BV z\|)\\\left(1+
\frac 1{2\theta}(\BV w-\BV v)\cdot\delta\BV U(\BV x,\BV x+\BV z)+
O(\|\delta\BV U\|^2)\right)f(\BV x,\BV v)f(\BV x+\BV z,\BV w)\dif\BV
z\dif\BV w.
\end{multline}
Therefore, we separate the collision integral in \eqref{eq:collision2}
into three parts:
\begin{subequations}
\label{eq:collision_split}
\begin{equation}
\coll[f]=\coll_1[f]+\coll_2[f]+\coll_3[f],
\end{equation}
\begin{equation}
\coll_1[f]=-\frac 1m f(\BV x,\BV v)\int_{\RR^3}\int_{B(\sigma)}
\parderiv{\phi(\|\BV z\|)}{\BV z}e^{ -\frac{\phi(\|\BV z\|)}\theta}
Y(\|\BV z\|)f(\BV x+\BV z,\BV w)\dif\BV z\dif\BV w,
\end{equation}
\begin{equation}
\coll_2[f]=-\frac{f(\BV x,\BV v)}{2m\theta}\int_{\RR^3}\int_{B(\sigma)
}\parderiv{\phi(\|\BV z\|)}{\BV z}e^{-\frac{\phi(\|\BV z\|)}\theta}
Y(\|\BV z\|)(\BV w-\BV v)\cdot\delta\BV Uf(\BV x+\BV z,\BV w)\dif\BV
z\dif \BV w,
\end{equation}
\begin{equation}
\coll_3[f]=-\frac 1mf(\BV x,\BV v)\int_{\RR^3}\int_{B(\sigma)}
\parderiv{\phi(\|\BV z\|)}{\BV z}e^{-\frac{\phi(\|\BV z\|)}\theta}
Y(\|\BV z\|)O(\|\delta\BV U\|^2)f(\BV x+\BV z,\BV w)\dif\BV z\dif\BV w.
\end{equation}
\end{subequations}
Henceforth we refer to $\coll_1[f]$ as the ``standard'' part of the
collision integral, since it is already present in our works
\cite{Abr22,Abr23,Abr24,Abr26,Abr27,Abr25}. Subsequently, $\coll_2[f]$
is dubbed the ``novel'' part. We will also show that $\coll_3[f]$
vanishes in the constant-density hydrodynamic limit \cite{Abr17}.

\subsection{Computation of the standard part of the collision integral}
\label{sec:standard_collision}

The collision integral $\coll_1[f]$ in \eqref{eq:collision_split} has
already been computed in our past works (see, for example,
Appendix~B.1 of \cite{Abr24}). Here we present the derivation of
$\coll_1[f]$ for the sake of completeness. First, we recall the
definition of density $\rho$ from \eqref{eq:rho_u_p}, so that
$\coll_1[f]$ can be expressed via
\begin{equation}
\coll_1[f]=-\frac 1mf(\BV x,\BV v)\int_{B(\sigma)}\parderiv{\phi(\|\BV
  z\|)}{\BV z}e^{-\frac{\phi(\|\BV z\|)}\theta}Y(\|\BV z\|)\rho(\BV x+
\BV z)\dif\BV z.
\end{equation}
Next, recalling that $\phi(r)$ has the effective range $\sigma$, we
thus denote
\begin{equation}
\label{eq:rescaled_phi}
\phi(r)=\tilde\phi(r/\sigma),\qquad Y(r)=\tilde Y(r/\sigma),
\end{equation}
where $\tilde\phi(r)$ has a unit range. Upon rescaling the dummy
variable of integration $\BV z\to \sigma\BV z$, we arrive at
\begin{equation}
\coll_1[f]=-\frac{\sigma^2}m f(\BV x,\BV v)\int_{B(1)}\parderiv{\tilde
  \phi(\|\BV z\|)}{\BV z}e^{-\frac{\tilde\phi(\|\BV z\|)}\theta}\tilde
Y(\|\BV z\|)\rho(\BV x+\sigma\BV z)\dif\BV z.
\end{equation}
The next step is to evaluate $\coll_1[f]$ in the hydrodynamic limit,
that is, as $\sigma\to 0$. Here we assume that the mass density
$\rho(\BV x+\sigma\BV z)$ is smooth in its argument, and thus we can
expand it in powers of $\sigma$. It is easy to see that the leading
order term of the expansion (that is, for $\rho(\BV x+\sigma\BV
z)\to\rho(\BV x)$) integrates to zero, and, therefore, we need to
examine the higher order terms in $\sigma$. In this case, however, we
can no longer assume that the temperature $\theta$ is a constant, and,
due to \eqref{eq:Y}, $\tilde Y$ is no longer a function of solely the
interparticle distance $r$:
\begin{equation}
\theta=\theta(\BV x),\qquad\tilde Y=\tilde Y(\BV x,r).
\end{equation}
Due to the symmetry reasons, we evaluate $\theta$ and $\tilde Y$ at
the midpoint $\BV x+\sigma\BV z/2$ between the coordinates of the
particles:
\begin{equation}
\label{eq:thetaY_midpoint}
\coll_1[f]=-\frac{\sigma^2}m f(\BV x,\BV v)\int_{B(1)}\parderiv{\tilde
  \phi(\|\BV z\|)}{\BV z}e^{-\frac{\tilde\phi(\|\BV z\|)}{\theta(\BV x
    +\sigma\BV z/2)}}\tilde Y(\BV x+\sigma\BV z/2,\|\BV z\|)\rho(\BV x
+\sigma\BV z)\dif\BV z.
\end{equation}
Here, we first observe that
\begin{subequations}
\begin{multline}
\parderiv{}{\BV z}\left(1- e^{-\frac{\tilde\phi(\|\BV z\|)}{\theta(\BV
    x+\sigma\BV z/2)}}\right)=e^{-\frac{\tilde\phi(\|\BV z\|)}{\theta(
    \BV x+\sigma\BV z/2)}}\parderiv{}{\BV z}\left(\frac{\tilde\phi(\|
  \BV z\|)}{\theta(\BV x+\sigma\BV z/2)}\right)\\=e^{-\frac{\tilde
    \phi(\|\BV z\|)}{\theta(\BV x+\sigma\BV z/2)}}\left(\frac 1{\theta
  (\BV x+\sigma\BV z/2)}\parderiv{\tilde\phi(\|\BV z\|)}{\BV z}-
\frac{\tilde\phi(\|\BV z\|)}{\theta^2(\BV x+\sigma\BV z/2)}\parderiv{
  \theta(\BV x+\sigma\BV z/2)}{\BV z}\right)\\=\frac{e^{-\frac{\tilde
      \phi(\|\BV z\|)}{\theta(\BV x+\sigma\BV z/2)}}}{\theta(\BV x+
  \sigma\BV z/2)}\left(\parderiv{\tilde\phi(\|\BV z\|)}{\BV z}-\frac
\sigma 2\frac{\tilde\phi(\|\BV z\|)}{\theta(\BV x+\sigma\BV z/2)}
\parderiv{\theta(\BV x+\sigma\BV z/2)}{\BV x}\right),
\end{multline}
\begin{multline}
\parderiv{}{\BV x}\left(1-e^{-\frac{\tilde\phi(\|\BV z\|)}{\theta(\BV
    x+\sigma\BV z/2)}}\right)=e^{-\frac{\tilde\phi(\|\BV z\|)}{\theta(
    \BV x+\sigma\BV z/2)}}\parderiv{}{\BV x}\left(\frac{\tilde\phi(\|
  \BV z\|)}{\theta(\BV x+\sigma\BV z/2)}\right)\\=-\frac{e^{-\frac{
      \tilde\phi(\|\BV z\|)}{\theta(\BV x+\sigma\BV z/2)}}}{\theta(\BV
  x+\sigma\BV z/2)}\frac{\tilde\phi(\|\BV z\|)}{\theta(\BV x+\sigma\BV
  z/2)}\parderiv{\theta(\BV x+\sigma\BV z/2)}{\BV x},
\end{multline}
\end{subequations}
and, therefore,
\begin{equation}
\parderiv{\tilde\phi(\|\BV z\|)}{\BV z}e^{-\frac{\tilde\phi(\|\BV z\|)
  }{\theta(\BV x+\sigma\BV z/2)}}=\theta(\BV x+\sigma\BV z/2)\left(
\parderiv{}{\BV z}-\frac\sigma 2\parderiv{}{\BV x}\right)\left(1-
e^{-\frac{\tilde\phi(\|\BV z\|)}{\theta(\BV x+\sigma\BV z/2)}}\right).
\end{equation}
This, in turn, leads to
\begin{multline}
-\int_{B(1)}\parderiv{\tilde\phi(\|\BV z\|)}{\BV z}e^{-\frac{\tilde\phi(r)}{
    \theta(\BV x+\sigma\BV z/2)}}\tilde Y(\BV x+\sigma\BV z/2,\|\BV z
\|)\rho(\BV x+\sigma\BV z)\dif\BV z\\=-\int_{B(1)}\theta(\BV x+\sigma\BV z/2)
\tilde Y(\BV x+\sigma\BV z/2,\|\BV z\|)\rho(\BV x+\sigma\BV z)
\left(\parderiv{}{\BV z}-\frac\sigma 2\parderiv{}{\BV x}\right)\left(
1-e^{-\frac{\tilde\phi(\|\BV z\|)}{\theta(\BV x+\sigma\BV z/2)}}
\right)\dif\BV z\\=\frac\sigma 2\parderiv{}{\BV x}\int_{B(1)}\left(1-e^{-
  \frac{\tilde\phi(\|\BV z\|)}{\theta(\BV x+\sigma\BV z/2)}}\right)
\theta(\BV x+\sigma\BV z/2)\tilde Y(\BV x+\sigma\BV z/2,\|\BV z\|)
\rho(\BV x+\sigma\BV z)\dif\BV z\\+\int_{B(1)}\left(1-e^{-\frac{\tilde\phi(
    \|\BV z\|)}{\theta(\BV x+\sigma\BV z/2)}}\right)\left(\parderiv{}{
  \BV z}-\frac\sigma 2\parderiv{}{\BV x}\right)\left[\theta(\BV x+
  \sigma\BV z/2)\tilde Y(\BV x+\sigma\BV z/2,\|\BV z\|)\rho(\BV x+
  \sigma\BV z)\right]\dif\BV z.
\end{multline}
Next, we observe that
\begin{subequations}
\begin{multline}
\left(\parderiv{}{\BV z}-\frac\sigma 2\parderiv{}{\BV x}\right)\left[
  \theta(\BV x+\sigma\BV z/2)\tilde Y(\BV x+\sigma\BV z/2,\|\BV z\|)
  \rho(\BV x+\sigma\BV z)\right]\\=\theta(\BV x+\sigma\BV z/2)\left(
\frac\sigma 2\tilde Y(\BV x+\sigma\BV z/2,\|\BV z\|)\parderiv{\rho(\BV
  x+\sigma\BV z)}{\BV x}+\rho(\BV x+\sigma\BV z)\parderiv{}r\tilde
Y(\BV x+\sigma\BV z/2,\|\BV z\|)\frac{\BV z}{\|\BV z\|}\right),
\end{multline}
\begin{multline}
\parderiv{}{\BV x}\left[\left(1-e^{-\frac{\tilde\phi(\|\BV z\|)}{
      \theta(\BV x+\sigma\BV z/2)}}\right)\theta(\BV x+\sigma\BV z/2)
  \tilde Y(\BV x+\sigma\BV z/2,\|\BV z\|)\rho(\BV x+\sigma\BV z)
  \right]\\+\theta(\BV x+\sigma\BV z/2)\tilde Y(\BV x+\sigma\BV z/2,
\|\BV z\|)\parderiv{\rho(\BV x+\sigma\BV z)}{\BV x}\\=\frac 1{\rho(\BV
  x+\sigma\BV z)}\parderiv{}{\BV x}\left[\left(1-e^{-\frac{\tilde\phi(
      \|\BV z\|)}{\theta(\BV x+\sigma\BV z/2)}}\right)\theta(\BV x+
  \sigma\BV z/2)\tilde Y(\BV x+\sigma\BV z/2,\|\BV z\|)\rho^2(\BV x+
  \sigma\BV z)\right].
\end{multline}
\end{subequations}
Thus, the integral becomes
\begin{multline}
-\int_{B(1)}\parderiv{\tilde\phi(\|\BV z\|)}{\BV z}e^{-\frac{\tilde\phi(r)}{
    \theta(\BV x+\sigma\BV z/2)}}\tilde Y(\BV x+\sigma\BV z/2,\|\BV z
\|)\rho(\BV x+\sigma\BV z)\dif\BV z\\=\frac\sigma 2\int_{B(1)}\frac 1{\rho(
  \BV x+\sigma\BV z)}\parderiv{}{\BV x}\left[\left(1-e^{-\frac{\tilde
      \phi(\|\BV z\|)}{\theta(\BV x+\sigma\BV z/2)}}\right)\tilde Y(
  \BV x+\sigma\BV z/2,\|\BV z\|)\rho^2(\BV x+\sigma\BV z)\theta(\BV x+
  \sigma\BV z/2)\right]\dif\BV z\\+\int_{B(1)}\left(1-e^{-\frac{\tilde\phi(\|
    \BV z\|)}{\theta(\BV x+\sigma\BV z/2)}}\right)\parderiv{}r\tilde
Y(\BV x+\sigma\BV z/2,\|\BV z\|)\frac{\BV z}{\|\BV z\|}\rho(\BV x+
\sigma\BV z)\theta(\BV x+\sigma\BV z/2)\dif\BV z.
\end{multline}
In the first sub-integral, the leading order term in $\sigma$ is
obtained by setting $\sigma=0$ everywhere inside the integral, i.e.
\begin{multline}
\frac\sigma 2\int_{B(1)}\frac 1{\rho(\BV x+\sigma\BV z)}\parderiv{}{\BV x}
\left[\left(1-e^{-\frac{\tilde\phi(\|\BV z\|)}{\theta(\BV x+\sigma\BV
      z/2)}}\right)\tilde Y(\BV x+\sigma\BV z/2,\|\BV z\|)\rho^2(\BV
  x+\sigma\BV z)\theta(\BV x+\sigma\BV z/2)\right]\dif\BV z\\=\frac
\sigma 2\frac 1{\rho(\BV x)}\parderiv{}{\BV x}\left[\rho^2(\BV x)
  \theta(\BV x)\int_{B(1)}\left(1-e^{-\frac{\tilde\phi(\|\BV z\|)}{\theta(\BV
      x)}}\right)\tilde Y(\BV x,\|\BV z\|)\dif\BV z\right]+
O(\sigma^2).
\end{multline}
In the second sub-integral, such a leading order term is zero, and thus
we need to differentiate in $\sigma$:
\begin{multline}
\int_{B(1)}\left(1-e^{-\frac{\tilde\phi(\|\BV z\|)}{\theta(\BV x+\sigma\BV
    z/2)}}\right)\parderiv{}r\tilde Y(\BV x+\sigma\BV z/2,\|\BV z\|)
\frac{\BV z}{\|\BV z\|}\rho(\BV x+\sigma\BV z)\theta(\BV x+\sigma\BV
z/2)\dif\BV z\\=\sigma\parderiv{}\sigma\!\int_{B(1)}\!\!\!\left.\left(
1-e^{-\frac{ \tilde\phi(\|\BV z\|)}{\theta(\BV x+\sigma\BV z/2)}}\right)\!
\parderiv{}r\tilde Y(\BV x+\sigma\BV z/2,\|\BV z\|)\frac{\BV z}{\|\BV
  z\|}\rho(\BV x+\sigma\BV z)\theta(\BV x+\sigma\BV z/2)\dif\BV z
\right |_{\sigma=0}\!\!\!\!\!\!\!\!\!+O(\sigma^2).
\end{multline}
In turn, we observe that
\begin{multline}
\parderiv{}\sigma\left[\left(1-e^{-\frac{\tilde\phi(\|\BV z\|)}{
      \theta(\BV x+\sigma\BV z/2)}}\right)\parderiv{}r\tilde Y(\BV x+
  \sigma\BV z/2,\|\BV z\|)\frac{\BV z}{\|\BV z\|}\rho(\BV x+\sigma\BV
  z)\theta(\BV x+\sigma\BV z/2)\right]\\=\rho(\BV x+\sigma\BV z)
\parderiv{}\sigma\left[\left(1-e^{-\frac{\tilde\phi(\|\BV z\|)}{\theta
      (\BV x+\sigma\BV z/2)}}\right)\parderiv{}r\tilde Y(\BV x+\sigma
  \BV z/2,\|\BV z\|)\frac{\BV z}{\|\BV z\|}\theta(\BV x+\sigma\BV z/2)
  \right]\\+\left(1-e^{-\frac{\tilde\phi(\|\BV z\|)}{\theta(\BV x+
    \sigma\BV z/2)}}\right)\parderiv{}r\tilde Y(\BV x+\sigma\BV z/2,
\|\BV z\|)\frac{\BV z}{\|\BV z\|}\theta(\BV x+\sigma\BV z/2)\parderiv{
}\sigma\rho(\BV x+\sigma\BV z)\\=\frac 12\rho(\BV x+\sigma\BV z)\frac{
  \BV z^2}{\|\BV z\|}\parderiv{}{\BV x}\left[\left(1-e^{-\frac{\tilde
      \phi(\|\BV z\|)}{\theta(\BV x+\sigma\BV z/2)}}\right)\parderiv{}
  r\tilde Y(\BV x+\sigma\BV z/2,\|\BV z\|)\theta(\BV x+\sigma\BV z/2)
  \right]\\+\left(1-e^{-\frac{\tilde\phi(\|\BV z\|)}{\theta(\BV x+
    \sigma\BV z/2)}}\right)\parderiv{}r\tilde Y(\BV x+\sigma\BV z/2,
\|\BV z\|)\theta(\BV x+\sigma\BV z/2)\frac{\BV z^2}{\|\BV z\|}
\parderiv{}{\BV x}\rho(\BV x+\sigma\BV z)\\=\frac 12\frac 1{\rho(\BV x
  +\sigma\BV z)}\frac{\BV z^2}{\|\BV z\|}\parderiv{}{\BV x}\left[
  \left(1-e^{-\frac{\tilde\phi(\|\BV z\|)}{\theta(\BV x+\sigma\BV z/2)
  }}\right)\parderiv{}r\tilde Y(\BV x+\sigma\BV z/2,\|\BV z\|)
  \rho^2(\BV x+\sigma\BV z)\theta(\BV x+\sigma\BV z/2)\right],
\end{multline}
and thus
\begin{multline}
\int_{B(1)}\left(1-e^{-\frac{\tilde\phi(\|\BV z\|)}{\theta(\BV x+\sigma\BV
    z/2)}}\right)\parderiv{}r\tilde Y(\BV x+\sigma\BV z/2,\|\BV z\|)
\frac{\BV z}{\|\BV z\|}\rho(\BV x+\sigma\BV z)\theta(\BV x+\sigma\BV
z/2)\dif\BV z\\=\frac\sigma 2\frac 1{\rho(\BV x)}\parderiv{}{\BV x}
\cdot\left[\rho^2(\BV x)\theta(\BV x)\int_{B(1)}\left(1-e^{-\frac{\tilde\phi(
      \|\BV z\|)}{\theta(\BV x)}}\right)\parderiv{}r\tilde Y(\BV x,\|
  \BV z\|)\frac{\BV z^2}{\|\BV z\|}\dif\BV z\right]+O(\sigma^2).
\end{multline}
We, therefore, arrive at
\begin{multline}
-\int_{B(1)}\parderiv{\tilde\phi(\|\BV z\|)}{\BV z}e^{-\frac{\tilde\phi(r)}{
    \theta(\BV x+\sigma\BV z/2)}}\tilde Y(\BV x+\sigma\BV z/2,\|\BV z
\|)\rho(\BV x+\sigma\BV z)\dif\BV z=\frac\sigma{2\rho(\BV x
  )}\parderiv{}{\BV x}\cdot\bigg[\rho^2(\BV x)\theta(\BV x)\\\int_{B(1)}\left(
  1-e^{-\frac{\tilde\phi(\|\BV z\|)}{\theta(\BV x)}}\right)\left(
  \tilde Y(\BV x,\|\BV z\|)\BM I+\parderiv{}r\tilde Y(\BV x,\|\BV z
  \|)\frac{\BV z^2}{\|\BV z\|}\right)\dif\BV z\bigg]+O(\sigma^2).
\end{multline}
Next, we switch to the spherical coordinates: $\BV z=r\BV n$, $\dif\BV
z=r^2\dif r\dif\BV n$, where $\BV n$ is a vector on the unit sphere
$\mathbb S_1$. In the spherical coordinates, the integral above
becomes
\begin{multline}
-\int_{B(1)}\parderiv{\tilde\phi(\|\BV z\|)}{\BV z}e^{-\frac{\tilde\phi(r)}{
    \theta(\BV x+\sigma\BV z/2)}}\tilde Y(\BV x+\sigma\BV z/2,\|\BV z
\|)\rho(\BV x+\sigma \BV z)\dif\BV z=\frac\sigma 2\frac 1{\rho(\BV x
  )}\\\parderiv{}{\BV x}\cdot\left[\rho^2(\BV x)\theta(\BV x)\int_{\mathbb S_1}
  \int_0^1\left(
  1-e^{-\frac{\tilde\phi(r)}{\theta(\BV x)}}\right)\left(\tilde Y(\BV
  x,r)\BM I+r\parderiv{}r\tilde Y(\BV x,r)\BV n^2\right)r^2\dif
  r\dif\BV n\right]+O(\sigma^2).
\end{multline}
The integrals over the angles are
\begin{equation}
\label{eq:intdn}
\int_{\mathbb S_1}\dif\BV n=4\pi,\qquad\int_{\mathbb S_1}\BV n^2\dif
\BV n=\frac{4\pi}3\BM I,
\end{equation}
which further yields
\begin{multline}
-\int_{B(1)}\parderiv{\tilde\phi(\|\BV z\|)}{\BV z}e^{-\frac{\tilde\phi(r)}{
    \theta(\BV x+\sigma\BV z/2)}}\tilde Y(\BV x+\sigma\BV z/2,\|\BV z
\|)\rho(\BV x+\sigma \BV z)\dif\BV z\\=\frac{2\pi\sigma}{\rho(\BV x)}
\parderiv{}{\BV x}\left[\rho^2(\BV x)\theta(\BV x)\int_0^1\left(
  1-e^{-\frac{\tilde\phi(r)}{\theta(\BV x)}}\right)\left(\tilde Y(\BV
  x,r)+\frac r3\parderiv{}r\tilde Y(\BV x,r)\right)r^2\dif r\right]+
O(\sigma^2)\\=\frac{2\pi\sigma}{3\rho(\BV x)}\parderiv{}{\BV x}\left[
  \rho^2(\BV x)\theta(\BV x)\int_0^1\left(1-e^{-\frac{\tilde\phi(
      r)}{\theta(\BV x)}}\right)\parderiv{}r\left(r^3\tilde Y(\BV x,r)
  \right)\dif r\right]+ O(\sigma^2).
\end{multline}
Subsequently, the collision integral $\coll_1[f]$ in
\eqref{eq:collision_split} is given via
\begin{equation}
\coll_1[f]=\frac{2\pi}3\frac{\sigma^3}m\frac{f(\BV x,\BV v)}{\rho(\BV
  x)}\parderiv{}{\BV x}\left[\rho^2(\BV x)\theta(\BV x)\int_0^1
  \left(1-e^{-\frac{\tilde\phi(r)}{\theta(\BV x)}}\right)\parderiv{}r
  \left(r^3\tilde Y(\BV x,r)\right)\dif r\right]+O(\sigma^4/m).
\end{equation}
As we can see, in the hydrodynamic limit $\sigma\to 0$, the
leading-order term of the collision integral remains finite and does
not vanish as long as $\sigma^3/m\sim$ const, that is, the molecular
density remains finite in the hydrodynamic limit \cite{Abr17}.
Discarding the higher-order term above and reverting back to $\phi$
and $Y$, we obtain
\begin{equation}
\coll_1[f]=\frac{2\pi}{3m}\frac{f(\BV x,\BV v)}{\rho(\BV x)}\parderiv{
}{\BV x}\left[\rho^2(\BV x)\theta(\BV x)\int_0^\sigma\left(1-e^{-\frac
    {\phi(r)}{\theta(\BV x)}}\right)\parderiv{}r\left(r^3 Y(\BV x,r)
  \right)\dif r\right],
\end{equation}
which translates directly into the first term in \eqref{eq:collision},
with the mean field potential given via \eqref{eq:bphi}.

\subsection{Computation of the novel part of the collision integral}
\label{sec:novel_collision}

To evaluate the novel part of the collision integral in
\eqref{eq:collision_split} in the hydrodynamic limit, we use the same
substitution as in \eqref{eq:rescaled_phi}, and change the variable
$\BV z\to\sigma\BV z$, which leads to
\begin{multline}
\coll_2[f]=-\frac{\sigma^2f(\BV x,\BV v)}{2m\theta}\int_{\RR^3}\int_{
  B(1)}\parderiv{\tilde\phi(\|\BV z\|)}{\BV z}e^{-\frac{\tilde\phi(\|
    \BV z\|)}{\theta(\BV x+\sigma\BV z/2)}}\tilde Y(\BV x+\sigma\BV z/
2,\|\BV z\|)\\(\BV w-\BV v)\cdot\delta\BV U(\BV x,\BV x+\sigma\BV z)
f(\BV x+\sigma\BV z,\BV w) \dif\BV z\dif\BV w.
\end{multline}
Following \eqref{eq:thetaY_midpoint}, here $\theta$ and $\tilde Y$ are
also evaluated at the midpoint between the coordinates of interacting
particles. Here, we assume that
\begin{equation}
|\delta\BV U(\BV x,\BV x+\sigma\BV z)|=|\BV U(\BV x+\sigma\BV z,\BV
x)-\BV U(\BV x,\BV x+\sigma\BV z)|=O(\sigma).
\end{equation}
This assumption relies on the fact that, unlike the density, the
velocity (and temperature, for that matter) does not have to possess
the ``potential wells'' around overlapping particle states, the Gibbs
equilibrium state being a prominent example of that.

In the constant-density hydrodynamic limit \cite{Abr17}, $\sigma\to 0$
with $\sigma^3/m\sim$ const. Thus, we expand $\theta$, $\tilde Y$ and
$f$ in powers of $\sigma$, and switch to spherical coordinates, which
leads to
\begin{subequations}
\begin{equation}
\coll_2[f]=\BM J\int_{\RR^3}(\BV w-\BV v)f(\BV x,\BV v)f(\BV x,\BV
w)\dif\BV w+O(\sigma^4/m),
\end{equation}
\begin{equation}
\BM J=-\frac{\sigma^3}{2m\theta}\int_0^1e^{-\frac{\tilde\phi(r)}{
    \theta(\BV x)}}\tilde Y(\BV x,r)\tilde\phi'(r)\left(\int_{\mathbb
  S_1}\BV n\frac{\delta\BV U(\BV x,\BV x+\sigma r\BV n)^T}\sigma \dif
\BV n\right)r^2\dif r.
\end{equation}
\end{subequations}
We subsequently discard the higher-order effect $O(\sigma^4/m)$ in the
hydrodynamic limit, and revert back to the original variables
$\phi(r)$ and $Y(r)$, which leads to the second term in
\eqref{eq:collision}, with $\BM J$ given in \eqref{eq:J}. Following
the same procedure with $\coll_3[f]$ in \eqref{eq:collision_split}, we
see that the whole integral is $O(\sigma^4/m)$, and thereby vanishes
in the constant-density hydrodynamic limit.

\section{Computation of the attenuation coefficient}
\label{sec:attenuation}

Here, we take the system \eqref{eq:rho_u_p_HS_divU} as a starting
point.  First, we discard the van der Waals effect and the collision
integrals from \eqref{eq:rho_u_p_HS_divU}, and substitute
\eqref{eq:Newton_Fourier} for $\BV\Sigma$ and $\BV q$, obtaining
\begin{subequations}
\begin{equation}
\Dderiv\rho t+\rho\nabla\cdot\BV u=0,\qquad\rho\Dderiv{\BV u}t+\nabla
p=\mu\left(\Delta\BV u+\frac 13\nabla(\nabla\cdot\BV u)\right),
\end{equation}
\begin{equation}
\Dderiv pt+\frac 53 p\nabla\cdot\BV u=\frac 23\mu\left(\nabla\BV u+
\nabla\BV u^T-\frac 23(\nabla\cdot\BV u)\BM I\right):\nabla\BV u\\+
\frac 23\frac\kappa R\Delta\theta+\frac 83\frac{\alpha\sigma_{SB}} {
  R^4}(\theta_0^4-\theta^4),
\end{equation}
\end{subequations}
where we used the kinetic temperature $\theta=RT$ in the expression
for the heat flux term. The above system represents the usual
compressible Navier--Stokes equations for a monatomic gas, with the
additional radiative cooling effect.

Next, we linearize the above system around the
background state $\rho=\rho_0$, $\BV u=\BV 0$, $\theta=\theta_0$, and
$p=p_0=\rho_0\theta_0$, with $\rho'$, $\theta'$ and $p'$ being small
perturbations:
\begin{subequations}
\begin{equation}
\parderiv{\rho'}t+\rho_0\nabla\cdot\BV u=0,\qquad\rho_0\parderiv{\BV
  u}t+\nabla p'=\mu\left(\Delta\BV u+\frac 13\nabla(\nabla\cdot\BV u)
\right),
\end{equation}
\begin{equation}
\Dderiv{p'}t+\frac 53 p_0\nabla\cdot\BV u=\frac 23\frac\kappa R\Delta
\theta'-\frac{32}3\frac{\alpha\sigma_{SB}T_0^3}R\theta'.
\end{equation}
\end{subequations}
Here, observe that, first, the density and pressure equations depend
only on $\nabla\cdot\BV u$, and second, it is more convenient to
switch to the $\theta'$-variable from the $p'$-variable. Therefore, we
compute the divergence of the linearized momentum equation above, and
switch to $\theta'$. The result is
\begin{subequations}
\begin{equation}
\parderiv{\rho'}t+\rho_0\chi=0,\qquad\parderiv\chi t+\frac{\theta_0}{
  \rho_0}\Delta\rho'+\Delta\theta'=\frac 43\frac\mu{\rho_0}\Delta\chi,
\end{equation}
\begin{equation}
\parderiv{\theta'}t+\frac 23\theta_0\chi=\frac 23\frac \kappa{R\rho_0}
\Delta\theta'-\frac{32}3\frac{\alpha\sigma_{SB}T_0^4}{p_0}\theta',
\end{equation}
\end{subequations}
where we use a separate notation $\chi=\nabla\cdot\BV u$, for
convenience. Next, we switch to the following nondimensional
variables:
\begin{equation}
\BV{\tilde x}=\frac{\BV x}L,\qquad\tilde t=\frac t{t_0},\qquad
\tilde\rho'=\frac{\rho'}{\rho_0},\qquad\tilde\chi=t_0\chi,\qquad
\tilde\theta'=\frac{\theta'}{\theta_0},
\end{equation}
where $L$ and $t_0$ are the spatial and temporal scales, to be chosen
below as necessary. In the nondimensional variables, the linearized
system becomes
\begin{subequations}
\begin{equation}
\parderiv{\tilde\rho'}{\tilde t}+\tilde\chi=0,\qquad\parderiv{\tilde
  \chi}{\tilde t}+\theta_0\frac{t_0^2}{L^2}\tilde\Delta(\tilde\rho'+
\tilde\theta')=\frac 43\frac{\mu t_0}{\rho_0L^2}\tilde\Delta\tilde
\chi,
\end{equation}
\begin{equation}
\parderiv{\tilde\theta'}{\tilde t}+\frac 23\tilde\chi=\frac 23\frac{
  \kappa t_0}{R\rho_0L^2}\tilde\Delta\tilde\theta'-\frac{32}3\frac{
  \alpha\sigma_{SB}T_0^4t_0}{p_0}\tilde\theta'.
\end{equation}
\end{subequations}
Now, we choose $t_0$ and $L$ as
\begin{equation}
t_0=\frac 3{32}\frac{p_0}{\alpha\sigma_{SB}T_0^4},\qquad L=\varepsilon
t_0 \sqrt{\frac 53\theta_0},\qquad\varepsilon=\left(\frac\mu{p_0t_0}
\right)^{1/2}=\left(\frac{32}3\frac{\alpha\sigma_{SB}T_0^4\mu}{p_0^2}
\right)^{1/2},
\end{equation}
where we introduced a small parameter $\varepsilon$, for
convenience. For the values of parameters above given in
Table~\ref{tab:ref_param}, we compute
\begin{equation}
t_0\approx 4.7\text{ days},\qquad\varepsilon\approx 2.1\cdot 10^{-8},
\qquad L\approx 3.2\text{ meters}.
\end{equation}
The equations become
\begin{equation}
\parderiv{\tilde\rho'}{\tilde t}+\tilde\chi=0,\qquad\parderiv{\tilde
  \chi}{\tilde t}+\frac 3{5\varepsilon^2}\tilde\Delta(\tilde\rho'+
\tilde\theta')=\frac 45\tilde\Delta\tilde\chi,\qquad\parderiv{\tilde
  \theta'}{\tilde t}+\frac 23\tilde\chi=\frac 1\Pran\tilde\Delta
\tilde\theta'-\tilde\theta',
\end{equation}
where the definition of the Prandtl number $\Pran$ is given in
\eqref{eq:attenuation}. In the Fourier space, the system of PDE above
becomes the system of ODE:
\begin{equation}
\deriv{\hat\rho'}{\tilde t}=-\tilde\chi,\qquad\deriv{\hat\chi}{\tilde
  t}=\frac 35\frac{\|\BV k\|^2}{\varepsilon^2}(\hat\rho'+\hat\theta')-
\frac 45\|\BV k\|^2\tilde\chi,\qquad\deriv{\hat\theta'}{\tilde t}=-
\frac 23\hat\chi-\left(1+\frac{\|\BV k\|^2}\Pran\right)
\hat\theta'.
\end{equation}
The matrix of the system is
\begin{equation}
\BM A=\begin{pmatrix} 0 & -1 & 0 \\ \frac 35\varepsilon^{-2}\|\BV k
\|^2 & -\frac 45\|\BV k\|^2 & \frac 35\varepsilon^{-2}\|\BV k\|^2 \\ 0
& -\frac 23 & -1-\frac 1\Pran\|\BV k\|^2 \end{pmatrix}.
\end{equation}
The characteristic equation is
\begin{equation}
\lambda^3+\left[1+\left(\frac 45+\frac 1\Pran\right)\|\BV k\|^2\right]
\lambda^2+\left[1+\frac{4\varepsilon^2}5\left(1+\frac{\|\BV k\|^2}
  \Pran\right)\right]\frac{\|\BV k\|^2}{\varepsilon^2}\lambda+\frac{3
  \|\BV k\|^2}{5\varepsilon^2}\left(1+\frac{\|\BV k\|^2}\Pran\right)=0
.
\end{equation}
This is a cubic equation, which means that we will have to use the
Cardano formula to compute the roots. For convenience, we denote
\begin{equation}
B=1+\left(\frac 45+\frac 1\Pran\right)\|\BV k\|^2,
\end{equation}
and make the substitution
\begin{equation}
\tilde\lambda=\lambda+\frac B3.
\end{equation}
This leads to
\begin{subequations}
\begin{equation}
\tilde\lambda^3+P\tilde\lambda+Q=0,\qquad P=\frac{\|\BV k\|^2}{
  \varepsilon^2}\left\{1+\varepsilon^2\left[\frac 45\left(1+\frac{
    \|\BV k\|^2}\Pran\right)-\frac{B^2}{3\|\BV k\|^2}\right]\right\},
\end{equation}
\begin{equation}
Q=\frac{\|\BV k\|^2}{\varepsilon^2}\left\{\frac 35\left(1+\frac{\|\BV
  k\|^2}\Pran\right)-\frac B3+\varepsilon^2\frac{2B}3\left[\frac{B^2}{
    9\|\BV k\|^2}-\frac 25\left(1+\frac{\|\BV k\|^2}\Pran\right)
  \right]\right\}.
\end{equation}
\end{subequations}
It is clear that the cubic discriminant
\begin{equation}
D=\left(\frac Q2\right)^2+\left(\frac P3\right)^3>0,
\end{equation}
which means that we have one real root $\tilde\lambda_0$, and a
complex-conjugate pair $\tilde\lambda_{1,2}$, given by the Cardano
formula:
\begin{equation}
\tilde\lambda_0=\xi_--\xi_+,\qquad\tilde\lambda_{1,2}=\frac{1\pm i
  \sqrt 3}2\xi_+-\frac{1\mp i\sqrt 3}2\xi_-,\qquad\xi_\pm=\left(\sqrt
D\pm\frac Q2\right)^{1/3}.
\end{equation}
Here, we will assume that $\varepsilon^{-1}\gg\|\BV
k\|\gg\varepsilon$, since $\|\BV k\|\sim 1$ corresponds to the spatial
scale of meters. If so, then we can use $\varepsilon$ as a small
parameter, and express
\begin{subequations}
\begin{equation}
P^3=\frac{\|\BV k\|^6}{\varepsilon^6}\left(1+O(\varepsilon^2)\right),
\qquad Q^2=O(\varepsilon^{-4}),\qquad D=\left(\frac{\|\BV k\|^2}{3
  \varepsilon^2}\right)^3\left(1+O(\varepsilon^2)\right),
\end{equation}
\begin{equation}
\sqrt D=\left(\frac{\|\BV k\|^2}{3\varepsilon^2}\right)^{3/2}\left(1+
O(\varepsilon^2)\right), \qquad Q=\frac{\|\BV k\|^2}{\varepsilon^2}
\left[\frac 35\left(1+\frac{\|\BV k\|^2}\Pran\right)-\frac B3+
  O(\varepsilon^2)\right],
\end{equation}
\begin{equation}
\sqrt D\pm\frac Q2=\left(\frac{\|\BV k\|^2}{3\varepsilon^2}\right)^{3/
  2}\left(1\pm\frac{3^{3/2}}2\frac\varepsilon{\|\BV k\|}\left[\frac 35
  \left(1+\frac{\|\BV k\|^2}\Pran\right)-\frac B3\right]+
O(\varepsilon^2)\right),
\end{equation}
\begin{equation}
\xi_\pm=\left(\sqrt D\pm\frac Q2\right)^{1/3}=\frac{\|\BV k\|}{\sqrt 3
  \varepsilon}\pm\frac 12\left[\frac 35 \left(1+\frac{\|\BV k\|^2}
  \Pran\right)-\frac B3\right]+O(\varepsilon),
\end{equation}
\begin{equation}
\tilde\lambda_0=-\frac 35 \left(1+\frac{\|\BV k\|^2}\Pran\right)+\frac
B3+O(\varepsilon),\qquad\tilde\lambda_{1,2}=\frac 12\left[\frac 35
  \left(1+\frac{\|\BV k\|^2} \Pran\right)-\frac B3\right]\pm i\frac{
  \|\BV k\|}\varepsilon+O(\varepsilon),
\end{equation}
\begin{equation}
\lambda_0=-\frac 35\left(1+\frac{\|\BV k\|^2}\Pran\right)
+O(\varepsilon), \qquad\lambda_{1,2}=-\frac 15\left[1
+\left(2+\frac 1\Pran\right)\|\BV k\|^2\right]\pm i\frac{
  \|\BV k\|}\varepsilon+O(\varepsilon).
\end{equation}
\end{subequations}
Observe that, as long as $\Pran<1$ (which is the case for common
gases), the real part of $\lambda_{1,2}$ is greater than $\lambda_0$
for all $\|\BV k\|$, and thereby constitutes the requisite attenuation
coefficient. Switching back to the dimensional variables, for the
attenuation coefficient in the physical space we obtain
\begin{equation}
a=\frac 1{5t_0}\left[1-\left(2+\frac 1\Pran\right)L^2\Delta\right]=
\frac 1{3\rho}\left[\frac{32}5\frac{\alpha\sigma_{SB}T_0^3}R-\left(2+
  \frac 1\Pran\right)\mu\Delta\right],
\end{equation}
which is the same expression as in \eqref{eq:attenuation}.

\section{Linear analysis}

\subsection{Linear wave structure}
\label{sec:waves}

The matrix of the system \eqref{eq:nondim_waves} is
\begin{equation}
\BM A=\begin{pmatrix}0 & -1 & 0 \\ \frac{3\beta\varepsilon^2}{20}\|\BV
k\|^2 & -\frac 45\varepsilon^2\|\BV k\|^2 & \frac{3\beta\varepsilon^2
}{80\eta}\|\BV k\|^2 \\ \varepsilon^2\left(\beta+
\frac{\|\BV k\|^2}\Pran\right) & -\frac 53 & -\frac{\|\BV k\|^2}{1+\|
    \BV k\|^2}-\varepsilon^2\left(\beta+\frac{\|\BV k\|^2}\Pran\right)
\end{pmatrix}.
\end{equation}
Here, for a known eigenvalue $\lambda$ of the matrix $\BM A$ above,
the corresponding eigenvector is
\begin{equation}
\BV e_\lambda=\begin{pmatrix}\frac 3{5\lambda}(\lambda+\frac{\|\BV k
  \|^2}{1+\|\BV k\|^2})z+O(\varepsilon^2/\lambda)\\ -\frac 35(\lambda
+\frac{\|\BV k\|^2}{1+\|\BV k\|^2})z+O(\varepsilon^2)\\ 1 \end{pmatrix}.
\end{equation}
The characteristic equation is
\begin{multline}
\lambda^3+\left[\frac{\|\BV k\|^2}{1+\|\BV k\|^2}+\varepsilon^2
  \left(\beta+\frac 45\|\BV k\|^2+ \frac{\|\BV k\|^2}\Pran\right)
  \right]\lambda^2\\+ \varepsilon^2\|\BV k\|^2\left[\frac{\beta
  }{16\eta}+\frac{3\beta}{20}+\frac 45\frac{\|\BV k\|^2}{1+\|\BV
    k\|^2}+\frac 45\varepsilon^2 \left(\beta+ \frac{\|\BV
    k\|^2}\Pran\right)\right]\lambda\\+\frac{3\beta\varepsilon^2}{20}\|\BV
k\|^2\left[\frac{\|\BV k\|^2}{1+\|\BV k\|^2}+\varepsilon^2
  \left(1+\frac 1{4\eta}\right)\left(\beta+ \frac{\|\BV
    k\|^2}\Pran\right)\right]=0.
\end{multline}
The leading order terms are dominant as long as $\varepsilon
\eta^{-1/2}\ll\|\BV k\|\ll\eta^{1/2}\varepsilon^{-1}$, or
$10^{-5}\ll\|\BV k\|\ll 10^5$, which corresponds to the scale range
between fractions of a millimeter and hundreds of kilometers.
Therefore, we simplify the characteristic equation as
\begin{equation}
\lambda^3+(B+\varepsilon^2B')\lambda^2+\frac{\beta\varepsilon^2 \|\BV
  k\|^2}{16\eta}\lambda+\frac{3\beta\varepsilon^2\|\BV k\|^2}{20}B=0,
\end{equation}
\begin{equation}
B=\frac{\|\BV k\|^2}{1+\|\BV k\|^2},\qquad B'=\beta+\frac 45\|\BV
k\|^2+ \frac{\|\BV k\|^2}\Pran.
\end{equation}
\begin{equation}
\tilde\lambda=\lambda+\frac 13(B+\varepsilon^2B'),
\end{equation}
and obtain the depressed cubic equation:
\begin{subequations}
\begin{equation}
\tilde\lambda^3+P\tilde\lambda+Q=0,\qquad P=-\frac{B^2}3+\varepsilon^2
\left(\frac{\beta\|\BV k\|^2}{16\eta}-\frac{2BB'}3\right)
+O(\varepsilon^4),
\end{equation}
\begin{equation}
Q=\frac 2{27}B^3+\varepsilon^2B\left(\frac{3\beta\|\BV k\|^2}{20}-
\frac 13\frac{\beta\|\BV k\|^2}{16\eta}+\frac 29BB'\right)+
O(\varepsilon^4).
\end{equation}
\end{subequations}
The discriminant is
\begin{multline}
D=\left(\frac Q2\right)^2+\left(\frac P3\right)^3=\frac{B^6}{27^2}+
\frac 1{27}\varepsilon^2B^4\left(\frac{3\beta\|\BV k\|^2}{20}-\frac 13
\frac{\beta\|\BV k\|^2}{16\eta}+\frac 29BB'\right)\\-\frac{B^6}{9^3}
+\frac{B^4}{9^2}\varepsilon^2 \left(\frac{\beta\|\BV k\|^2}{16\eta}-
\frac{2BB'}3\right)+O(\varepsilon^4)=\frac{\varepsilon^2B^4\beta\|\BV
  k\|^2}{180}+O(\varepsilon^4)>0,
\end{multline}
and thus the Cardano formula applies:
\begin{equation}
\tilde\lambda_0=\xi_--\xi_+,\qquad\tilde\lambda_{1,2}=\frac{1\pm i
  \sqrt 3}2\xi_+-\frac{1\mp i\sqrt 3}2\xi_-,\qquad\xi_\pm=\left(\sqrt
D\pm\frac Q2\right)^{1/3}.
\end{equation}
We subsequently have
\begin{subequations}
\begin{equation}
\sqrt D\pm\frac Q2=\pm\frac{B^3}{27}+\frac{\varepsilon B^2\sqrt\beta
  \|\BV k\|}{6\sqrt 5}\pm\frac{\varepsilon^2B}2\left(\frac{3\beta\|\BV
  k\|^2}{20}-\frac 13\frac{\beta\|\BV k\|^2}{16\eta}+\frac
29BB'\right)+O(\varepsilon^3),
\end{equation}
\begin{equation}
\xi_\pm=\pm\frac B3+\frac{\varepsilon\sqrt\beta\|\BV k\|}{2\sqrt 5}\pm
\frac{\varepsilon^2}{2B}\left(\frac{3\beta\|\BV k\|^2}{20}-\frac{\beta
  \|\BV k\|^2}{16\eta}+\frac 23BB'\right)+O(\varepsilon^3),
\end{equation}
\begin{equation}
\tilde\lambda_0=-\frac{2B}3+O(\varepsilon^2),\quad\tilde\lambda_{1,2}
=\frac B3\pm\frac {i\varepsilon\|\BV k\|}2\sqrt{\frac{3\beta}5}+\frac{
  \varepsilon^2}{2B}\left(\frac{3\beta\|\BV k\|^2}{20}-\frac{\beta
  \|\BV k\|^2}{16\eta}+\frac 23BB'\right)+O(\varepsilon^3),
\end{equation}
\begin{equation}
\lambda_0=-B+O(\varepsilon^2),\qquad\lambda_{1,2}=\pm\frac{i\varepsilon
  \|\BV k\|}2\sqrt{\frac{3\beta}5} +\frac{\varepsilon^2}{2B}\left(
\frac{3\beta\|\BV k\|^2}{20}-\frac{\beta\|\BV k\|^2}{16\eta}+\frac
23BB'\right)+O(\varepsilon^3).
\end{equation}
\end{subequations}
Ignoring the $O(\eta)$-correction in the $O(\varepsilon^2)$-term of
$\lambda_{1,2}$, and reverting to original notations, we arrive at
\eqref{eq:lambda_waves}. Here, we have
\begin{equation}
\BV e_0=\begin{pmatrix} 0 \\ 0 \\ 1 \end{pmatrix}+O(\varepsilon^2),\qquad
\BV e_{1,2}=\begin{pmatrix} 1 \\ 0 \\ 0 \end{pmatrix}+O(\varepsilon),
\end{equation}
i.e., $\lambda_0$ is associated with a rapidly decaying pressure, and
$\lambda_{1,2}$ are slowly decaying density waves.

\subsection{Large scales}
\label{sec:large_scale}

For convenience, we introduce the following temporary notations:
\begin{subequations}
\begin{equation}
a=\frac{\|\BV k\|^2}{\Rey_\zeta}\sim 1,\qquad b=\frac{2k_x^2}{\|\BV k
  \|^2}\sim 1,\qquad c=\frac{2k_xk_y}{\|\BV k\|^2}-\frac{\vartheta\Rey
}{\Mach^2}\sim 1,
\end{equation}
\begin{equation}
d=\frac{\beta\Rey_\zeta^2}{4\Rey\Mach^2}\sim 1,\qquad\varepsilon=
\frac{\Rey_\zeta}\Rey\sim 10^{-3}.
\end{equation}
\end{subequations}
In the notations above, $\BM A$ in \eqref{eq:Fourier_space} is given
via
\begin{equation}
\BM A=\begin{pmatrix} 0 & 0 & -1 \\ 0 & -\varepsilon a & -1 \\
ad & b & c-a \end{pmatrix}.
\end{equation}
The characteristic equation is given via
\begin{equation}
\lambda^3+[(1+\varepsilon) a-c]\lambda^2+[b+ad+\varepsilon a(a-c)]
\lambda+\varepsilon a^2d=0.
\end{equation}
Although the above is a cubic equation, the roots can be evaluated
with a necessary accuracy without resorting to the cubic formulas,
thanks to the presence of a small parameter $\varepsilon$. First, note
that the free term is $\sim\varepsilon$, which means that one of the
roots is of the same magnitude. Substituting
$\lambda=\varepsilon\tilde\lambda$, we obtain
\begin{equation}
\tilde\lambda_0=-\frac a{1+\frac b{ad}}+O(\varepsilon),\qquad
\lambda_0=-\frac{\varepsilon a}{1+\frac b{ad}}+O(\varepsilon^2),
\end{equation}
which is \eqref{eq:l0_ls} in the original notations. We assume that
the remaining eigenvalues are $\sim 1$, which leads to the quadratic
equation
\begin{equation}
\lambda^2+(a-c)\lambda+b+ad+O(\varepsilon)=0.
\end{equation}
Depending on the balance of the coefficients, the roots can be either
real or comprise a complex-conjugate pair. We first investigate the
scenario with the real roots, for which the requirement is
\begin{equation}
\label{eq:discriminant}
(c-a)^2>4(b+ad).
\end{equation}
In this case, the roots are given via
\begin{equation}
\lambda_{1,2}=\frac{c-a}2\pm\frac{|c-a|}2\sqrt{1-4\frac{b+ad}{(c-a)^2}
}+O(\varepsilon).
\end{equation}
Since the square root is $<1$, the requirement for a positive root is
$c>a$. In this case, from \eqref{eq:discriminant} it follows that
\begin{equation}
c^2>4b,\qquad\text{or}\qquad\frac{4k_x^2k_y^2}{\|\BV k\|^4}>\frac{
  8k_x^2}{\|\BV k\|^2},\qquad\text{or}\qquad k_y^2>2\|\BV k\|^2.
\end{equation}
Since the latter condition never holds, we conclude that a positive
real part can only be achieved in a complex-conjugate pair of roots.
In such a case we have
\begin{equation}
\lambda_{1,2}=\frac{c-a}2\pm i\sqrt{b+ad-\frac{(c-a)^2}4}
+O(\varepsilon),
\end{equation}
which is \eqref{eq:l12_ls} in the original notations.

\subsection{Small scales}
\label{sec:small_scale}

For convenience, we introduce the following temporary notations:
\begin{equation}
a=\frac{\|\BV k\|^2}\Rey\sim 1,\qquad b=\frac{2k_x^2}{\|\BV k\|^2}\sim
1,\qquad c=\frac{2k_xk_y}{\|\BV k\|^2}-\frac{\vartheta\Rey}{\Mach^2}\sim 1,
\end{equation}
\begin{equation}
d=\frac{\beta\Rey_\zeta^2}{4\Rey\Mach^2}\sim 1,\qquad\varepsilon=\frac{
  \Rey_\zeta}\Rey\sim 10^{-3}.
\end{equation}
In the notations above, the matrix $\BM A$ in \eqref{eq:Fourier_space}
is given via
\begin{equation}
\BM A=\begin{pmatrix} 0 & 0 & -1 \\ 0 & -a & -1 \\
\varepsilon^{-1}ad & b & c-\varepsilon^{-1}a
\end{pmatrix}.
\end{equation}
The characteristic equation is given via
\begin{equation}
\varepsilon\lambda^3+[a+\varepsilon(a-c)]\lambda^2+[a(a+d)+\varepsilon
  (b-ac)]\lambda+a^2d=0.
\end{equation}
Here, one of the roots is $O(\varepsilon^{-1})$, and two are
$O(1)$. For the latter, the characteristic equation becomes
\begin{equation}
(\lambda+d)(\lambda+a)+O(\varepsilon)=0.
\end{equation}
The roots are obviously
\begin{equation}
\lambda_0=-d+O(\varepsilon),\qquad\lambda_1=-a+O(\varepsilon).
\end{equation}
For the $O(\varepsilon^{-1})$-root, we substitute
$\lambda=\varepsilon^{-1}\tilde\lambda$, and obtain
\begin{equation}
\tilde\lambda^3+a\tilde\lambda^2+O(\varepsilon)=0,\qquad\text{or}\qquad
\lambda_2=-\frac a\varepsilon+O(1).
\end{equation}
In the original notations, the expressions for the roots above become
\eqref{eq:l_ss}.

\subsection{Asymptotic behavior}
\label{sec:asymptotic}

In order to analyze the asymptotic behavior of
\eqref{eq:Fourier_space}, we make a change of the time variable as
\begin{equation}
\label{eq:tau}
 \tau(\tilde t)=\frac{k_x^2\tilde t^3}3+k_xk_{y,0}\tilde t^2+\|\BV k_0
 \|^2\tilde t=\tilde t\bigg[\frac{k_x^2\tilde t^2}{12}+\left(\frac 12
   k_x\tilde t+k_{y,0}\right)^2+k_x^2\bigg].
\end{equation}
Above, the expression in square parentheses is strictly greater than
zero for $\BV k_0\neq\BV 0$, and thus $\tau(\tilde t)$ is invertible
everywhere on the real line. The new time variable is chosen so that
$\tau'(\tilde t)=\|\BV k(\tilde t)\|^2$, which converts
\eqref{eq:Fourier_space} into
\begin{equation}
\deriv{}\tau\begin{pmatrix}\hat\rho \\ \hat\omega \\ \hat\chi
\end{pmatrix}=\BM B(\tau)\begin{pmatrix}\hat\rho \\ \hat\omega \\
\hat\chi\end{pmatrix},\qquad\BM B=\begin{pmatrix} 0 & 0 & -\frac
1{\|\BV k(\tilde t(\tau))\|^2} \\ 0 & -\frac 1\Rey & -\frac 1{\|\BV
  k(\tilde t(\tau))\|^2} \\ \frac{\beta\Rey_\zeta}{4\Rey\Mach^2} &
\frac{2k_x^2}{\|\BV k(\tilde t(\tau))\|^4} & \frac{2k_xk_y(\tilde
  t(\tau))}{\|\BV k(\tilde t(\tau))\|^4}-\frac{\vartheta\Rey}{
\Mach^2\|\BV k(\tilde t(\tau))\|^2}-\frac 1{\Rey_\zeta}
\end{pmatrix}.
\end{equation}
Here, we note that the system above is asymptotically autonomous
(because the time-dependent entries vanish as $\tau\to\infty$), with
$\BM B(\infty)$ having three distinct eigenvalues
\begin{equation}
\lambda_0=0,\qquad\lambda_1=-\frac 1\Rey,\qquad\lambda_2=-\frac 1{
  \Rey_\zeta},
\end{equation}
with the corresponding eigenvectors
\begin{equation}
\BV e_0(\infty)=\begin{pmatrix}1 \\ 0 \\ \frac{\beta\Rey_\zeta^2}{4\Rey
  \Mach^2} \end{pmatrix},\qquad\BV e_1(\infty)=\begin{pmatrix} 0 \\ 1
\\ 0\end{pmatrix},\qquad\BV e_2(\infty)=\begin{pmatrix} 0 \\ 0
\\ 1\end{pmatrix}.
\end{equation}
Therefore, Levinson's theorem \cite[Theorem~8.1]{CodLev} applies
directly, and we can use it to investigate the asymptotic
behavior. Following \cite{Abr27}, we introduce the quantity
\begin{equation}
\kappa(\tilde t)=\|\BV k(\tilde t)\|^2,\qquad\kappa'(\tilde t)=2\BV
k(\tilde t)\cdot\BV k'(\tilde t)=2k_xk_y(\tilde t).
\end{equation}
With the new notation, we have
\begin{equation}
\BM B=\begin{pmatrix} 0 & 0 & -\frac 1\kappa\\ 0 & -\frac 1\Rey &
-\frac 1\kappa \\ \frac{\beta\Rey_\zeta}{4\Rey\Mach^2} & \frac{2k_x^2}{
  \kappa^2} & \frac{\kappa'}{\kappa^2}-\frac{\vartheta\Rey}{
\Mach^2\kappa}-\frac 1{\Rey_\zeta}
\end{pmatrix}.
\end{equation}
The characteristic equation is

\begin{multline}
\lambda^3+\left(\frac 1{\Rey_\zeta}+\frac 1\Rey+\frac{\vartheta\Rey}{
  \Mach^2\kappa}-\frac{\kappa'}{\kappa^2}\right)\lambda^2\\+\left(\frac
1{\Rey_\zeta\Rey}+\frac{\beta\Rey_\zeta}{4\Rey\Mach^2\kappa}+\frac
\vartheta{\Mach^2\kappa}-\frac{\kappa' }{\Rey\kappa^2}+\frac{2k_x^2}{
  \kappa^3}\right)\lambda+\frac{\beta\Rey_\zeta}{4\Rey^2\Mach^2\kappa}=0.
\end{multline}
One root is $O(\kappa^{-1})$. Substituting $\tilde\lambda=
\kappa\lambda$, we obtain
\begin{equation}
\tilde\lambda_0=-\frac{\beta\Rey_\zeta^2}{4\Rey\Mach^2}+O(\kappa^{-1}),
\qquad\lambda_0=-\frac{\beta\Rey_\zeta^2}{4\Rey\Mach^2}\frac 1\kappa+
O(\kappa^{-2}).
\end{equation}
The remaining roots are $O(1)$. For them, we have
\begin{equation}
\left(\lambda+\frac 1\Rey\right)\left(\lambda+\frac
1{\Rey_\zeta}\right)+O(\kappa^{-1})=0,
\end{equation}
which leads to
\begin{equation}
\lambda_1=-\frac 1\Rey+O(\kappa^{-1}),\qquad\lambda_2=-\frac
1{\Rey_\zeta}+O(\kappa^{-1}).
\end{equation}
Now we have, in the leading order,
\begin{subequations}
\begin{equation}
\int_{\tau_0}^\tau\lambda_0\dif\tau=-\frac{\beta\Rey_\zeta^2}{4\Rey
  \Mach^2} \int_{\tau_0}^\tau\frac{\dif\tau}\kappa=-\frac{\beta
  \Rey_\zeta^2}{4\Rey\Mach^2} \int_{\tilde t_0}^{\tilde t}\frac{\kappa
  \dif\tilde t}\kappa=-\frac{\beta \Rey_\zeta^2(\tilde t-\tilde t_0)
}{4\Rey\Mach^2},
\end{equation}
\begin{equation}
\int_{\tau_0}^\tau\lambda_1\dif\tau=-\frac{\tau-\tau_0}\Rey,\qquad
\int_{\tau_0}^\tau\lambda_2\dif\tau=-\frac{\tau-\tau_0}{\Rey_\zeta},
\end{equation}
\end{subequations}
and, therefore, according to Levinson's theorem, asymptotically we
have
\begin{equation}
\begin{pmatrix}\tilde\rho \\ \tilde\omega \\ \tilde\chi\end{pmatrix}
  \sim C_0e^{-\frac{\beta\Rey_\zeta^2\tilde t}{4\Rey\Mach^2}}\begin{pmatrix}
    1 \\ 0 \\ \frac{\beta\Rey_\zeta^2}{4\Rey\Mach^2} \end{pmatrix}+
  C_1e^{-\frac{\tau(\tilde t)}\Rey}\begin{pmatrix}0 \\ 1 \\ 0 \end{pmatrix}
  + C_2e^{-\frac{\tau(\tilde t)}{\Rey_\zeta}}\begin{pmatrix} 0 \\ 0 \\ 1
\end{pmatrix}.
\end{equation}
Substituting the expression for $\tau(\tilde t)$ from \eqref{eq:tau},
and using \eqref{eq:Re2}, we arrive at \eqref{eq:l_asymptotic}.

\end{document}